\documentstyle[12pt]{article}  %\def\emptypage{\thispagestyle{empty}~\newpage }
\let\3=\ss \catcode`\"=\active \let"=\"     \long\def\del#1\enddel{ }
  \oddsidemargin=\evensidemargin
 \let\msk=\medskip \let\bsk=\bigskip
  
   \def\ve{\vfil\eject}

\let\a=\alpha \let\b=\beta   \let\e=\varepsilon
  \let\th=\theta  
\let\l=\lambda    \let\p=\pi 
   \let\c=\chi

\def\0{\over }    \def\1{\vec }   \def\2{{1\over2}} \def\3{{\ss}}
\def\4{{1\over4}} \def\5{\bar }   \def\6{\partial } \def\7#1{{#1}\llap{/}}
\def\8#1{{\textstyle{#1}}}        \def\9#1{{\bf {#1}}}
\def\_#1{$\underline{\hbox{#1}}$} \def\^#1{$\overline{\hbox{#1}}$}

\def\<{\langle } \def\>{\rangle }  
 
\def \({\left( } \def \){\right) }

     \let\aus=\in
      \let\and=\wedge
     
\def\|#1{{}_{\bigg|_{#1}}}

\def\mao#1{\mathop{\rm #1}\nolimits}      \def\tr{\mao{tr}} 
  \def\mod{\mao{mod}}
 \def\gcd{\mao{gcd}} 
\ifx\Box  \else \fi

\def\pmbf#1{\setbox0=\hbox{${#1}$}   \kern-.025em\copy0\kern-\wd0
      \kern.05em\copy0\kern-\wd0     \kern-.025em\raise.0433em\box0 }
   
  \def\cg{g} 
 \def\cm{{\cal M}} \def\co{{\cal O}} 
\def\cx{{\cal X}} % Poor man's      BLACKBOARD BOLD     char often used:
\def\inbar{\vrule height1.5ex width.4pt depth0pt} %\font\ZZsf=cmss12
\def\ifundefined#1{\expandafter\ifx\csname#1\endcsname\relax}
\makeatletter \ifundefined{new@mathgroup} {} \else % \input{oldlfont.sty} \fi
\new@mathgroup\sffam
\new@internalmathalphabet\mathsf\sffam{cmss}{m}{n}
\def\psf{\fontfamily\sfdefault \fontseries\default@series
    \fontshape\default@shape\selectfont\mathsf}
\fi
\def\ZZ{\relax{\sf Z\kern-.4em \sf Z}}  \def\IR{\relax{\rm I\kern-.18em R}}
\def\IN{\relax{\rm I\kern-.18em N}} \def\IP{\relax{\rm I\kern-.18em P}}
\def\IQ{\relax\,\hbox{$\inbar\kern-.3em{\rm Q}$}}
\def\IC{\hbox{\,$\inbar\kern-.3em{\rm C}$}}
\def\citen#1{\if@filesw \immediate\write \@auxout {\string\citation{#1}}\fi%
\@tempcntb\m@ne \let\@h@ld\relax \def\@citea{}%
\@for \@citeb:=#1\do {\@ifundefined {b@\@citeb}%
    {\@h@ld\@citea\@tempcntb\m@ne{\bf ?}%
    \@warning {Citation `\@citeb ' on page \thepage \space undefined}}%
    {\@tempcnta\@tempcntb \advance\@tempcnta\@ne
    \setbox\z@\hbox\bgroup\ifcat0\csname b@\@citeb \endcsname \relax
       \egroup \@tempcntb\number\csname b@\@citeb \endcsname \relax
       \else \egroup \@tempcntb\m@ne \fi \ifnum\@tempcnta=\@tempcntb
       \ifx\@h@ld\relax \edef \@h@ld{\@citea\csname b@\@citeb\endcsname}%
       \else \edef\@h@ld{\hbox{--}\penalty\@highpenalty
	      \csname b@\@citeb\endcsname}\fi
    \else \@h@ld\@citea\csname b@\@citeb \endcsname \let\@h@ld\relax \fi}%
 \def\@citea{,\penalty\@highpenalty\hskip.13em plus.13em minus.13em}}\@h@ld}
\def\@citex[#1]#2{\@cite{\citen{#2}}{#1}}%
\def\@cite#1#2{\leavevmode\unskip
  \ifnum\lastpenalty=\z@\penalty\@highpenalty\fi% highpenalty before
  \ [{\multiply\@highpenalty 3 #1%              % triple-highpenalties within.
  \if@tempswa,\penalty\@highpenalty\ #2\fi}]}   % and before note.
\makeatother % END OF   "CITE.STY"   (\citen can be used for citation numbers)

\def\beq{\begin{equation}} \def\eeq{\end{equation}} \def\eql#1{\label{#1}\eeq}
\def\bea{\begin{eqnarray}} \def\eea{\end{eqnarray}} 
\def\fnote#1#2{\begingroup\def\thefootnote{#1}\footnote{#2}
	   \addtocounter{footnote}{-1}\endgroup}    

\def\plb#1 #2 {Phys. Lett. {\bf B#1} #2 }
\def\phr#1 #2 {Phys. Rep. {\bf  #1} #2 } 
\def\npb#1 #2 {Nucl. Phys. {\bf B#1} #2 }
\def\aph#1 #2 {Ann. Phys. {\bf #1} #2 }  
\def\jmp#1 #2 {J. Math. Phys. {\bf #1} #2 }
\def\prd#1 #2 {Phys. Rev. {\bf D#1} #2 }
\def\prl#1 #2 {Phys. Rev. Lett. {\bf #1} #2 }
\def\rmp#1 #2 {Rev. Mod. Phys.  {\bf #1} #2 }
\def\zpc#1 #2 {Z. Phys. {\bf #1C} #2 }
\def\cmp#1 #2 {Comm. Math. Phys. {\bf #1} #2 }
\def\mpl#1 #2 {Mod. Phys. Lett. {\bf A#1} #2 }
\def\ijmp#1 #2 {Int. J. Mod. Phys. {\bf A#1} #2 }

\ifundefined{emptypage} \let\emptypage=\relax \fi
\def\naive{na\"\i ve} \def\[{\left[} \def\]{\right]} \let\cdots=\ldots
\def\V{{\vrule height13pt width0pt depth7pt}}   \def\lbo{\linebreak[0]}
   \def\PP{Poincar\'e polynomial}
\def\ng{n_{27}} \def\na{n_{\overline{27}}} \let\ngb=\na
\def\tbf#1:{{\noindent\bf #1:}} \def\new#1\endnew{{\bf #1}}

\def\figuresonly{\pagestyle{empty}\figa\ve\figb\ve\figc\end{document}}
\long\def\old#1\endold{{\small #1}}         \def\oldansw{o } \def\cutansw{c }
\newcount{\npre} \npre=1  \def\negansw{s } \def\figansw{f } \def\textansw{t }
\def\ifpre{\ifnum\npre=1 } \def\ifsub{\ifnum\npre=0 }        \def\cut#1{#1}
\def\askversion{\message{
Preprint (p) / submit (s) / text only (t) / figures only (f):  (p/s/t/f)? }
    \read-1 to\answ \ifx\answ\negansw \npre=0 \else \npre=1 \fi
    \ifx\answ\figansw { } \else \def\figuresonly{ }   \fi
    \ifx\answ\oldansw \def\old##1\endold{{\small ##1}}\fi
    \ifx\answ\textansw \npre=2 \else \message{
Cut figures (use 'c' in case of memory problem):  (c/n)? }
    \read-1 to\answ\ifx\answ\cutansw \def\cut##1{}\npre=7\fi\fi \figuresonly }

\def\bpic{\begin{picture}} \def\epic{\end{picture}} \thicklines
\def\lab#1)#2#3{\put#1){\makebox(0,0)[#2]{\small #3}}}
\def\putlin#1,#2,#3,#4,#5){\put#1,#2){\line(#3,#4){#5}}} %\putlin(x,y,dx,dy,l)
\def\putvec#1,#2,#3,#4,#5){\put#1,#2){\vector(#3,#4){#5}}}

\newcount{\vmul} \newcount{\hdiv} % hor.length /=(\hdiv) vert.length *=(\vdiv)
\newcount{\wi}   \newcount{\he}   % width (height) of figure in \unitlength
\newcount{\hq}   \newcount{\vq}   % hor. (vert.) quantities
\newcount{\hoff} \newcount{\auxc} \newcounter{figco}   \def\npt{\circle*{2}}

\def\vlline{\put(-3,0){\line(1,0)6}}      % linemark for labels on vert. axis
\def\vlright#1{\put(6,0){\makebox(0,0)[l]{\scriptsize #1}}}
\def\putvl#1{\mbox{\bpic(0,0)\funit=1pt\vlline\vlright{#1}\epic}}
\def\putvm#1{\mbox{\bpic(0,0)\funit=1pt\vlline\epic}}   % linemark (no label)
\def\vlab#1{\vq=#1\multiply\vq by\vmul \put(-\hoff,\vq){\putvl{#1}}
	    \put(\hoff,\vq){\putvm{#1}} }
\def\vmark#1{\vq=#1\multiply\vq by\vmul \put(-\hoff,\vq){\putvm{#1}}
	    \put(\hoff,\vq){\putvm{#1}}  }

\def\hlline{\put(0,-3){\line(0,1)6}}     % linemark for labels on horiz. axis
\def\hltop#1{\put(0,6){\makebox(0,0)[b]{\scriptsize #1}}}
\def\puthl#1{\mbox{\bpic(0,0)\funit=1pt\hlline\hltop{#1}\epic}}
\def\hlab#1{\hq=#1\divide\hq by\hdiv \put(\hq,0){\puthl{#1}} }    % at origin:
\def\hlabo{\put(0,0){\mbox{\bpic(0,0)\funit=1pt\put(0,-3){\line(0,1)3}\epic}}}

\def\Vpt#1,#2){\hq=#2\advance\hq by -#1 \multiply\hq by 2 \divide\hq by\hdiv
	       \vq=#1\advance\vq by #2 \multiply\vq by\vmul\put(\hq,\vq){\npt}}
\def\Vplo#1{\vbox{\hdiv=2\vmul=1 \figsca \auxc=\he \multiply\auxc by\vmul
    \hoff=\wi\divide\hoff by2 \stepcounter{figco}\message{[Fig. \arabic{figco}}
    \begin{center}\let\.=\Vpt \bpic(\wi,\auxc)(-\hoff,0) \figlab #1 \hlabo
    \put(-\hoff,0){\framebox(\wi,\auxc){}} \epic \\[5mm]
    Fig. \arabic{figco}: \figcap \end{center}} \vfil \message{]}}
\def\figsca{\unitlength=1.1pt \wi=500 \he=400} \let\funit=\unitlength
\begin{document}      \def\hannover{ITP--UH--10/92} \def\cern{CERN-TH.6705/92}
{\hfill\cern\vskip-9pt \hfill\hannover\vskip-9pt    \hfill hep-th/9211047}
\vskip 15mm \centerline{\hss\large
       ALL ABELIAN SYMMETRIES OF LANDAU--GINZBURG POTENTIALS   \hss}
\begin{center} \vskip 8mm
       Maximilian KREUZER\fnote{*}{e-mail: kreuzer@cernvm.cern.ch} %\star
\vskip 3mm
       CERN, Theory Division\\
       CH--1211 Geneva 23, SWITZERLAND
\vskip 6mm               and
\vskip 3mm
       Harald SKARKE\fnote{\#}{e-mail: skarke@kastor.itp.uni-hannover.de}
\vskip 3mm
       Institut f"ur Theoretische Physik, Universit"at Hannover\\
       Appelstra\3e 2, D--3000 Hannover 1, GERMANY

\vfil                        {\bf ABSTRACT}                \end{center}

We present an algorithm for determining all inequivalent abelian symmetries of
non-degenerate quasi-homogeneous polynomials and apply it to the recently
constructed complete set of Landau--Ginzburg potentials for $N=2$
superconformal
field theories with $c=9$. A complete calculation of the resulting orbifolds
without torsion increases the number of known spectra by about one third. The
mirror symmetry of these spectra, however, remains at the same low level as for
untwisted Landau--Ginzburg models. This happens in spite of the fact that the
subclass of potentials for which the Berglund--H\"ubsch construction works
features perfect mirror symmetry. We also make first steps into the space of
orbifolds with $\ZZ_2$ torsions by including extra trivial fields.

\vfil\noindent \cern\\ \hannover\\ October 1992 \msk
\thispagestyle{empty} \newpage  \emptypage
\setcounter{page}{1} \pagestyle{plain}
\ifsub \baselineskip=20pt \else \baselineskip=14pt \fi

\section{Introduction}

Landau--Ginzburg (LG) models \cite{mvw,lvw} represent
a fairly general framework
for constructing $N=2$ superconformal field theories, which are needed for
supersymmetric string vacua \cite{N2}. They provide, for example, a link
between exactly solvable models and Calabi--Yau compactifications~\cite{gvw},
and also contain large classes of such models as special cases.
In general, LG theories cannot be solved exactly. Still, some basic
information on the massless spectrum of the resulting string models can be
extracted in a simple way from the superpotential owing to
non-renormalization properties. As a bonus, on the other hand, a
very efficient algorithm for the calculation of the
number of non-singlet representations of $E_6$ in (2,2) vacua can be
derived from the formulae for charge degeneracies of LG orbifolds given in
refs.~\cite{v,iv}. We therefore believe that it is worth while to invest some
effort into the classification of the string vacua that can be obtained in
this way.

Recently the basis for this work has been laid by the enumeration of all
(deformation classes of) non-degenerate potentials with central
charge $c=9$~\cite{nms,kle}. In the present paper, as a second step, we
calculate all abelian orbifolds that can result from manifest symmetries
at non-singular points in the moduli spaces of these potentials,
disregarding however the possibility of discrete torsions \cite{tor,fiqs}
(except for a modest probe into $\ZZ_2$ torsions). To do so, we %somewhat
extend the results of Vafa and
Intriligator~\cite{v,iv} for the calculation of the chiral ring in orbifolds
and prove an analogue of a well-known theorem for Calabi--Yau
manifolds~\cite{c} in the LG context.

Theoretically, our results are interesting for the question of mirror
symmetry~\cite{cls,gp,bh} and for the classification of $N=2$ models.
Unfortunately, for both of these (related) issues, our results are, in a
sense, negative, as even many spectra with a
large Euler number remain without mirror. From a phenomenological point of
view, however, extensions of our
calculations towards including torsion and non-abelian orbifolds look very
promising, because new spectra mainly arise in the realm of small particle
content in the effective field theory.

In section~2 we show how the (non-singlet) massless spectrum of an orbifold
can, in general, be obtained from the index and the dimension of the chiral
ring in a very efficient way. The fact that the computation of neither of
these quantities requires explicit knowledge of a basis for
the ring is vital for a complete investigation of all
non-degenerate cases. Only if there are states with left-right charges
$(q_L,q_R)=(1,0)$ or $(0,1)$ do we need extra information. We show how to
extract the number of such states and that they can exist only if Witten's
index vanishes.
Section~3 is quite technical and details how we construct, based upon
the classification of potentials \cite{cqf}, all possible (linear) abelian
symmetries. The reader who is only interested in the results may wish to
proceed to section~4. Implications of our findings are then
discussed in the final section.

\section{Hodge numbers}

In this section we review the results of refs. \cite{v,iv} and use them to
derive an efficient algorithm for the calculation of the spectrum in case of
vanishing torsions. We do, however, keep the discussion general as long as
possible.

The basic information about the massless spectrum of $(2,2)$ heterotic string
models
is contained in the chiral ring, i.e. the non-singular
operator product algebra of chiral primary fields \cite{gso}.
These fields have a linear relation between their conformal dimensions and
their $U(1)$ charges. Their charge degeneracies are conveniently summarized
by the \PP\, which, for left- and right-chiral fields, is defined as
\beq  P(t,\5t)=\tr_{(c,c)} t^{J_0}\5t^{\5J_0}. \eeq
Analogous generating functions can be defined for the charge degeneracies of
Ramond ground states and anti-chiral fields; they are related to one another
by spectral flow~\cite{lvw}.

For $N=2$ supersymmetric LG models, defined by a non-degenerate
quasi-homogeneous superpotential   \beq W(\l^{n_i}X_i)=\l^dW(X_i),   \eql{qh}
the \PP\ is given by \cite{bou}
\beq P(t,\5t)=P(t\5t)=\prod{1-(t\5t)^{1-q_i}\01-(t\5t)^{q_i}},        \eql{pp}
with $q_i=n_i/d$. Strictly speaking, this is a polynomial in $t^{1/d}$
and $\5t^{1/d}$ rather than in $t$ and $\5t$. In order to obtain a model
with integer charges we thus need to project onto states invariant under
the transformation
\beq j=e^{2\p iJ_0}=\ZZ_d[n_1,\ldots,n_N],                            \eql{zd}
which rotates the field $X_i$ by a phase $2\p iq_i$
in the complex plane (we will use the same symbol for the generator of a
cyclic group and for the group itself).
We can, of course, use any symmetry that contains this
projection to twist the original LG conformal field theory. Only under
certain conditions, however, will all the charges in the twisted sectors
of the resulting orbifold be integer and thus space-time supersymmetry be
unbroken~\cite{iv} (see below). We will restrict ourselves to models
where both left and right $U(1)$ charges are integer.

Even though some of these models definitely do not have an
interpretation in terms of a string propagating on a Calabi--Yau (CY)
manifold,\footnote{Namely those which would have $h_{11}=0$, in contradiction
	 with the requirement of the existence of a K\"ahler form.}
we will use CY terminology and call the coefficients of the \PP\
$P(t,\5t)=p_{ij}t^i\5t^i$, i.e. the numbers $p_{ij}$ of chiral primary fields
with charge $(q_L,q_R)=(i,j)$, Hodge numbers. Let $\ng$ and $\ngb$ denote
the numbers of 27 and
$\overline{27}$ representations of $E_6$ in the corresponding heterotic
string model. If we have a CY interpretation, then
$h_{11}=p_{12}=\ngb$, $h_{12}=p_{11}=\ng$, and the Euler number of the manifold
is $\c=2(h_{11}-h_{12})=2(p_{12}-p_{11})=2(\ngb-\ng)$.

In the twisted sectors of a LG orbifold only the fields that are
invariant under the twist should
contribute to Ramond ground states and to chiral primary fields. It can
then be shown \cite{v,iv} that the left/right charges of the Neveu--Schwarz
ground state
$|h\>$ in the sector twisted by an element $h$ of the symmetry group
are given by    \beq \sum_{\th_i^h>0}\; \8\2-q_i\pm(\th_i^h-\8\2),  \eql{job}
and that the action of a group element $g$ on that state is
\beq     g|h\>=(-1)^{K_gK_h}\e(g,h){\det g_{|_h}\0\det g}|h\>,       \eql{ga}
where $h$ is assumed to act diagonally with phases $0\le\th_i^h<1$ on the
fields $X_i$ and $g$ commutes with $h$. Here $\det g_{|_h}$ denotes the
determinant of the action of $g$ on the fields that are invariant under $h$.
There is a certain freedom in the phase of the action of $g$ on the
$h$-twisted sector, which is parametrized by the discrete torsions
$\e(g,h)$~\cite{tor} and by the integers $K_g$ mod~2 satisfying
$K_{gh}=K_g+K_h$ (${K_g}$ determines the sign of the action of $g$ in the
Ramond sector \cite{iv}).

For space-time supersymmetry, and hence integer left charges in the
internal conformal field theory, the symmetry group used for the modding must
contain the canonical $\ZZ_d$ symmetry (\ref{zd}) of the potential.
Furthermore, the state $|j^{-1}\>$, the analogue of the holomorphic 3-form,
should be invariant under the complete group, which fixes the torsions
$\e(j,g)=(-1)^{K_gK_j} \det g$ and $K_j=N-D$, where $N$ is the number of
fields and $D=c/3=\sum_{i\le N} (1-2q_i)$.  We will also demand
$(-1)^{K_g}=\det g$ to ensure that both left
and right charges are integer and that
the left-right symmetric spectral flow between the Neveu--Schwarz and the
Ramond sector is local \cite{iv}.

\subsection{Index and dimension of the chiral ring}

With the above information it is, in principle, straightforward to compute
the \PP\ for any given LG orbifold by summing over all invariant states for
all possible twisted sectors. To do so, however, we would explicitly need a
basis for the chiral ring, i.e. for the quotient of the polynomial ring by
the ring generated by the gradients $\6_iW$ of the potential. %~\cite{lvw}.
Fortunately, for $D=3$ there is a simpler approach, which allows us to treat
different types of potentials on an equal footing. In fact, in ref.~\cite{iv}
the transformation properties (\ref{ga}) of the twisted states have been
inferred from modular invariance of Witten's index $\tr_R\,(-1)^{J_0-\5J_0}$,
which is shown to be given by
\beq P(-1,-1)=-\c={1\0|G|}\sum_{gh=hg}(-1)^{N+K_gK_h+K_{gh}}\e(g,h)
     \prod_{\th_i^g=\th_i^h=0} {n_i-d\0n_i}.                           \eql x
As usual, this formula can be interpreted in two different ways: We can think
of it as the sum over the contributions with boundary conditions $g$ and $h$
in the space and time direction of the torus, respectively. The last
factor $\prod(d-n_i)/n_i$ in (\ref x) is just the dimension of the ring
restricted to fields invariant under $g$ and $h$,
as is obvious from formula (\ref{pp}) for the \PP. Alternatively, we may
consider the sum over $h$ to be the sum over twists, with the sum over $g$,
normalized by the dimension $|G|$ of the group, implementing the projection.

Using the latter interpretation, it is easy to obtain from (\ref x) a
formula for the sum of all entries in the Hodge diamond, which we denote by
$\bar\c=P(1,1)=4+2\ngb+2\ng+8p_{01}$
(for the second equality we have assumed $p_{01}=p_{02}=p_{10}=p_{20}$
and the Poincar\'e duality $P(t,\5t)=(t\5t)^DP(1/t,1/\5t)$).
As the chiral fields $X_i$ have left-right symmetric charges, all states
in a given twisted sector contribute with the same sign to the index (\ref x).
This sign can be computed from the ground state, for which, according to
(\ref{job}), $J_0-\5J_0=\sum_{\th_i^h>0}(2\th_i^h-1)$. As $\det h=(-1)^{K_h}$,
this is nothing but $K_h+N-N_h$ mod 2, where $N_h$ is the number of untwisted
fields in that sector. Correspondingly, $(-1)^{K_h+N-N_h}$ is the sign
of the contribution with $g=1$ to the index (\ref x). We thus obtain
\beq P(1,1)=\5\c={1\0|G|}\sum_{gh=hg}(-1)^{N_h+K_gK_h+K_{g}}\e(g,h)
     \prod_{\th_i^g=\th_i^h=0} {n_i-d\0n_i}                           \eql{xb}
for the dimension of the $(c,c)$-ring of the orbifold.

Given the analogue of the first Betti number, $p_1=p_{01}+p_{10}$,
we can use this information to compute $\ng$ and $\na$.
At first sight, eqs. (\ref x) and (\ref{xb})
seem to have the disadvantage that we need to go twice over the group,
implying that the computation time for an orbifold grows with the square
of the group order. In fact, however, this need not be the case if we restrict
ourselves to vanishing torsions. We will now show how this comes about, and
then determine  under what conditions $p_1$ need not vanish and how to
compute this number.

Our first simplification to this end is to assume $K_g=0$ for all $g$. This
is, in fact, no restriction, since we can always make a determinant positive by
including an additional trivial superfield $X_{N+1}$ with $q_{N+1}=1/2$,
and letting $g$ act non-trivially on that field. This modification does not
change the central charge of a theory, nor does it change the ring, because the
additional field can be eliminated by its equation of motion.
In particular, with $j$ contained in the group,
$\det j=1$ implies that for $D=3$ we require the number of fields to be odd.
With several trivial fields,
different actions of group elements $g$ and $h$ of even order on these fields
correspond to generically
different choices of $\ZZ_2$ torsions $\e(g,h)=\pm1$.

Restricting ourselves to $\e(g,h)=1$ for all $g$ and $h$, we obtain
\def\prodrat{\;\prod_{\th_i^g=\th_i^h=0}\;{q_i-1\0q_i}}  % {i:gX_i=hX_i=X_i}
\beq \c=|G|^{-1}\sum_{g\in G}\sum_{h\in G}\prodrat,                \eql{chi}
\beq \5\c=|G|^{-1}\sum_{g\in G}(-1)^{N_g}\;\sum_{h\in G}\prodrat,  \eql{bchi}
where $N_g$ denotes the number of $X_i$ that are invariant under $g$.
Special care has to be taken in their evaluation.
A \naive\ application would imply a number of operations proportional
to $|G|^2$.       \def\cx{{\{X_i\}}}
% Let $\cx$ denote the set of all $X_i$, $\cx=\{X_1,\cdots,X_N\}$.
If we assign to each subset $\cm\subset\cx$ of the set of fields
the number $n_\cm$ of group
elements that leave the elements of $\cm$ invariant while acting
non-trivially on all other $X_i$, we can rewrite these formulae as
\beq \c=|G|^{-1}\sum_{\cm\subset\cx}\sum_{\tilde\cm\subset\cx}
     n_\cm n_{\tilde\cm} \prod_{i:X_i\in\cm\cap\tilde\cm}{q_i-1\0q_i},    \eeq
\beq \c=|G|^{-1}\sum_{\cm\subset\cx}\sum_{\tilde\cm\subset\cx}(-1)^{|\cm|}
     n_\cm n_{\tilde\cm} \prod_{i:X_i\in\cm\cap\tilde\cm}{q_i-1\0q_i}.    \eeq
The advantage of this formulation is that we can avoid the double
summation over the group by first creating a list of the $n_\cm$'s (evaluating
each group element only once). Then the double summation takes place over
the subsets $\cm$ of $\cx$,
whose number is between 32 for $N=5$ and 512 for $N=9$, to be compared with
frequently occurring group orders of several thousands. Besides, we only have
to go over the $\tilde\cm$'s for those $\cm$ where $n_\cm\ne0$. There is a
natural bitwise representation for the sets $\cm$, namely setting the $i^{th}$
bit to 1 if $\cm$ contains $X_i$ and to 0 otherwise, and of course this bit
pattern can be identified with an integer. The operator $\&$ provided
by the programming language $C$ (bitwise logical and) then represents the
intersection of $\cm$ and $\tilde\cm$.

A fast algorithm using these tricks works the following way:\\
(1) Create an array of length $2^N$ containing the numbers $n_\cm$.\\
(2) Create arrays with entries $k_{\hat\cm}=\sum_{\cm\cap\tilde\cm=\hat\cm}
   n_\cm n_{\tilde\cm}$ and $\bar k_{\hat\cm}=\sum_{\cm\cap\tilde\cm=\hat\cm}
   (-1)^{|\cm|}n_\cm n_{\tilde\cm}$.\\
(3) Calculate
\beq \c=|G|^{-1}\sum_{\hat\cm\subset\cx}k_{\hat\cm}
	\prod_{i:X_i\in\hat\cm}{q_i-1\0q_i},    \eeq
\beq \bar\c=|G|^{-1}\sum_{\hat\cm\subset\cx}\bar k_{\hat\cm}
	\prod_{i:X_i\in\hat\cm}{q_i-1\0q_i},    \eeq
evaluating the time-consuming product over rational numbers only once for
each $\hat\cm$ with $(k_{\hat\cm}, \bar k_{\hat\cm})\ne (0,0)$.

\subsection{The first Betti number}

The final ingredient we need for the calculation of the Hodge diamond is the
number $p_{01}$.
In \cite{nms} we have shown that this number can be non-zero only if there
is a subset $\cm_1$ of the set of fields $\{X_i\}$ and an element $j_1$ of
the symmetry group such that $j_1$ acts like $j$ on the fields $X_i\aus\cm_1$
and does not act at all on the remaining fields. Furthermore, the contribution
of the fields in $\cm_1$ to the central charge has to be 3, i.e. $\sum_{X_i\aus
\cm_1}(1-2q_i)=1$. The proof of this statement in \cite{nms} applies without
modification to the general case with arbitrary torsions and $K_g$.
In addition, we have also shown there that, as for CY manifolds,
$p_1>0$ implies $\chi=0$, because any LG model with $p_1>0$
factorizes into the product of a torus times a conformal field theory with
$c=6$.
Although the factorization property does not generalize to the case of an
arbitrary orbifold, we will now show that $\chi=0$ can still be concluded,
so that we need to calculate $p_{01}$ only if the index vanishes.
   (Note that only twisted {\it vacua} can contribute to $p_{01}$; thus the
   calculation of this number does not require explicit knowledge of the ring
   either.)

In order to show this, we consider a state $|j_1\>$ contributing to $p_{01}$.
Of course, it has to be invariant under the whole group, implying
\beq(-1)^{K_{j_1}K_g}\e(g,j_1)={\det g\0\det g_{|_{j_1}}}
    =\det g_{|_{\cm_1}}. \eeq
In case of trivial torsions and $K_g=0$ this implies that any group element
must have determinant 1 on $\cm_1$.
We are going to show that, as a consequence, all twists $h$ that can contribute
to the \PP\ coincide with $(j_1)^p$, where $p\aus\{-1,0,1\}$, on $\cm_1$, which
in turn will imply that the \PP\ factorizes.
In particular, $p_{01}$ is the number of different subsets $\cm_1$ of $\cx$
that contribute 3 to the central charge,  for which there is a group element
that acts like $j$ on $\cm_1$ and trivially on the remaining fields, and with
all determinants of group elements equal to one on $\cm_1$. For $D=3$ this
number can only be 0, 1 or 3, with 3 corresponding to the 3-torus.

To prove the assertion,
consider a symmetry $g$ of $W(X_i)$ and let $\cx_g$ denote the set of
those $X_i$ that are themselves invariant under $g$. The restriction of $W$
to $\cx_g$, i.e. setting all other fields to 0, is also non-degenerate
(in the language of \cite{cqf}, no fields in $\cx_g$, and thus no links between
these fields can point out of that set).
\del Let us call the set $\IC_{(n_1,\ldots,n_N)}[d]$ of all non-degenerate
polynomials satisfying the quasi-homogeneity condition (\ref{qh}) a
configuration.\enddel
Then, with $g=j(j_1)^{-1}$, $W$ restricted to $\cm_1$ is a
non-degenerate potential with $D=1$.
\del There are, disregarding trivial
fields, only three such configurations, namely $\IC_{1,1,1}[3]$, $\IC_{1,1}[4]$
and $\IC_{1,2}[6]$. Due to our determinant condition, $W$ restricted to $\cm_1$
has to belong
to one of these configurations with an even or odd number of additional trivial
fields in the first or second and third case, respectively.

Now consider a twist $h$ that contributes chiral states to the orbifold
under investigation.
By considering all cases one can check that invariance under
$j_1$ implies that $h$ either acts trivially on
$\cm_1$ or non-trivially on all fields of $\cm_1$ (except, possibly, for a
trivial action on pairs of trivial fields). Another check of all possibilities
shows that this entails, with all determinants on $\cm_1$ equal to 1, that $h$
acts like $(j_1)^p$ on $\cm_1$ with $-1\le p\le1$, again with the trivial
exceptions.\enddel
Any twist $h$ can be decomposed as $h_1h_2$, where $h_1$ acts only on $\cm_1$
and $h_2$ acts only on the other fields.
Consider the $D=1$ torus twisted by $j_1$ and $h_1$.
The \PP\ of the $D=1$ torus is unique and all its entries come from
states twisted by powers $j_1^p$ of $j_1$ (unless there are at least two
trivial fields leading to a doubling of the ground state; this would not
affect our arguments, however).
Therefore the $h_1$-twisted states cannot
survive the $j_1$-projection unless $h_1$ is itself a power of $j_1$
($h_1$ does not project itself out).\footnote{These
   features cannot be generalized to orbifolds with
   torsion. It is straightforward to construct examples where twists with
   complex determinants on $\cm_1$ contribute to the chiral ring. Still, in
   all the examples we know, the conclusion $\chi=0$ is true.}
According to eq. (\ref{ga}) the action of $j_1$ on the $h_1$ twisted sector for
$D=1$ is the same as the action of $j_1$ on the $h$ twisted sector for $D=3$.
Thus the twists $h$ that contribute to $P(t,\5t)$ belong to
one of the following classes: Either $p=\pm1$, then the fields in $\cm_1$
contribute a factor $t$ or $\5t$, respectively, or $p=0$. In the latter
case the fields in $\cm_1$ contribute two states:
$q_L=q_R=0$ and $q_L=q_R=1$.
If any of these four contributions is present, then the other three,
with identical contributions from the remaining fields, will also occur.
Therefore the complete \PP\
has the form $P(t,\5t)=(1+t)(1+\5t)Q(t,\5t)$, and $\c=-P(-1,-1)=0$.

This factorization of the \PP\
 does {\it not} mean that the LG orbifold is a product of
a $c=3$ and a $c=6$ theory. For the latter there seem to be only two possible
spectra, namely those corresponding to the torus $T^2$ or the K3 surface,
whereas we find several different $c=9$ spectra with $p_{01}\ne 0$. The
reason is that not all states of a ``would-be product'' survive the
group projections. Consider, for example, the $1^9$ with the symmetries
$j_1=\ZZ_3[1,1,1,0,0,0,0,0,0]$,
$j_2=\ZZ_3[0,0,0,1,1,1,1,1,1]$ and $g=\ZZ_3[0,1,2,0,1,2,0,0,0]$. Without $g$,
the \PP\ would be
\beq P(t,\5t)=(1+t)(1+\5t)\((1+t^2)(1+\5t^2)+20t\5t\),          \eeq
where the $20t\5t$ stand for the $({6\atop 3})=20$ states $X_iX_jX_k|0\>$
with $4\le i,j,k\le9$ in the untwisted sector. The $g$ projection reduces
this number to $4+({4\atop 3})=8$, coming from states $X_4X_5X_i|0\>$ and
$X_iX_jX_k|0\>$, $6\le i,j,k\le9$.

\section{Finding the symmetries}

This section is based on the criterion for non-degeneracy of a configuration
$\IC_{(n_1,\ldots,n_N)}[d]$, i.e. the existence of a non-degenerate polynomial
that is quasi-homogeneous with respect to the weights $q_i=n_i/d$,
given in ref. \cite{cqf}. It is easy to see that for any such polynomial
there has to be a monomial of the form $X_i^{\a_i}$ or $X_i^{\a_i}X_j$ for
each field $X_i$. We call the second type of monomial a pointer from $X_i$
to $X_j$ and refer to the sum of $N$ such monomials as a skeleton for the
non-degenerate polynomial. Such a skeleton is in general not unique, and if
there is more than one pointer at the same field, additional monomials are
required for non-degeneracy~\cite{cqf}. We call the skeletons without such
double pointers invertible. They correspond to the polynomials that are
non-degenerate
with only $N$ monomials and play an important role in mirror symmetry.
Note that a skeleton already
determines a configuration and also fixes a maximal abelian phase symmetry,
which of course contains any phase symmetry of the complete potential.

The crucial simplification in considering {\it abelian} LG orbifolds is that
we can assume any abelian symmetry to be diagonalized, i.e. to act as a
phase symmetry. Thus we construct all possible inequivalent abelian symmetries
by first constructing the inequivalent skeletons of a configuration. Then
we determine the maximal phase symmetry of such a skeleton. Finally we
construct all subgroups of this symmetry that satisfy the relevant
set of conditions.

\subsection{All skeletons of a configuration}

Constructing all skeletons of a configuration is straightforward by choosing
all possible combinations of pointers: $X_i$ can point at $X_j$ iff $n_i$
divides $d-n_j$ (here it is convenient to say a
field points at itself iff it corresponds to a Fermat-type monomial $X_i^\a$;
these ``pointers'', however, do not count for the non-degeneracy criterion).
In the case of permutation symmetries of the configuration, i.e.
if some $n_i$ are equal, this procedure, however, can generate many equivalent
skeletons. It is convenient to represent each consistent
skeleton for a given configuration by an integer whose $i^{th}$ digit is the
index of the target of the $i^{th}$ field. This implies a natural ordering.
A simple concept for eliminating the redundancy is to compute for each
skeleton the set of all
permutations consistent with the quasi-homogeneity of the configuration.
Then we keep only those skeletons whose integer representation is minimal in
the respective set of equivalent skeletons.
This (admittedly crude) method requires only a few minutes to produce
106144 inequivalent skeletons for the 10838 non-degenerate configurations
with up to 8 fields, which were constructed in refs. \cite{nms,kle}.
For the unique configuration with 9 fields, however, which has
the maximal permutation symmetry, this method wastes about a day of
computing time. In that case we can alternatively use the algorithm of
ref.~\cite{nms} to compute all topologically inequivalent skeletons, which
takes about a second. As a check for our programs we have used both
algorithms to construct the 2615 different skeletons with 9 fields.

\subsection{The maximal abelian symmetry of a skeleton}

We will see that for a given skeleton the maximal abelian phase
symmetry is of order
$\co=\prod_i \a_i\;\prod_j\((\prod_{k\le l_j}\a_{jk})\,-\,(-1)^{l_j}\)$,
where the
first product extends over all exponents of fields that are not members
of a loop, while the second product extends over all loops $j$ with $l_j$
respective fields. This symmetry group can be represented by at most $N$
generators $\cg_i$, which we construct recursively.

We start with the fields that are not members of a loop and always
consider the origin of a pointer before its target. When arriving
at the field $Y$ with exponent $\b$, the typical situation is that
the $Y$-dependence of the skeleton  polynomial is given by
\beq ZY^\b+\sum_{i=1}^I YX_i^{\a_i}.                              \eeq
As an induction hypothesis we assume that for each such field $X_i$
there is, so far, only one generator $\cg_i$ under which $X_i$
transforms, and that $X_i$ transforms with a phase $1/\a_i$, i.e.
$\cg_iX_i=\exp(2\p i/\a)X_i$ (for convenience we omit the obvious
factor $2\p$ and thus have rational ``phases'').

Let $a_i$ be the order of $\cg_i$ and $U$ be the least
common multiple of the $a_i$. Choose a set of divisors $t_i$ of the $a_i$
such that $U=\prod t_i$ and $\gcd(t_i,t_j)=\gcd(t_i,a_i/t_i)=1$.
Given the prime decomposition $U=\prod p_j^{n_j}$ this means that each
number $p_j^{n_j}$ divides a particular $t_i$.
The group generated by $\cg_i$ is therefore the direct
product of  the groups generated by $\cg_i'=(\cg_i)^{t_i}$ and
by $(\cg_i)^{a_i/t_i}$.

Allowing now $Y$ to transform, but keeping $Z$ fixed, we obviously
enlarge the group order by the factor $\b$, because the $X_i^{\a_i}$ may
have $\b$ different phases under a group transformation.
We now construct a generator $\cg_Y$ of order $U\b$, which generates,
together with the $\cg_i'$ and all generators $\cg_k'$ already constructed,
the maximal phase symmetry that keeps $Z$ fixed. Let the phases of
$X_i$ and $Y$ under $\cg_Y$ be $-1/(\a_i\b)$ and $1/\b$, respectively. All
other fields, pointing at some $X_i$, transform with $-1/\b$ times
their phase under $\cg_i$.
Note that we never use a relation involving the group order to change a
phase to an equivalent one, so that all monomials are manifestly invariant,
even if we enlarge the group order by a factor. Thus, if a path of length
$n$ points from some field $X_k$ to $Y$, then the inverse phase
of $X_k$ under $\cg_Y$ is $(-1)^n$ times the product of all exponents
along the path. The order of $\cg_Y$ is $\b$ times $U$, which is equal
to the least common multiple of the products of the exponents along
maximal paths pointing at $Y$.
Furthermore, there are no non-trivial relations between $\cg_Y$ and the
$\cg_i'$, so that the group we have constructed has the correct order.

There is a slight complication if we eventually hit a field belonging
to a loop. Within a loop, the transformation of all fields is fixed by
the phase of any particular one. Thus,
for our purpose, a loop acts like a single
field. Let $Y_j$ point at $Y_{j+1}$ for $j<J$ and $Y_J$ point at
$Y_1$ with respective exponents $\b_j$. The maximal phase symmetry of the
loop is generated by $\cg_Y$ acting on $Y_1,Y_2,\ldots,Y_J$ with phases
$1/\co,-\b_1/\co,\b_1\b_2/\co,\ldots$, where $\co=\prod\b_i-(-1)^J$.
In the final step of the calculation of the maximal abelian symmetry of
a connected component of the skeleton we now have
to consider all fields $X_i$ pointing at the loop
and the respective generators $\cg_i$ of orders $a_i$. The generators $\cg_i'$
are constructed as above. If we used the same recipe as above to
construct the action of $\cg_Y$ on $X_i$ pointing at $Y_j$, $X_k$ pointing at
$X_i$, etc., we would define phases
\beq{-b_j\0\co}\({1\0\a_i},{-1\0\a_i\a_k},\ldots\),\qquad
                                       b_j=\prod_{l\le j}(-\b_l), \eeq
with $a_i$ being the least common denominator of the fractions in parenthesis.
Unfortunately, if $b_j$ and $a_i$ have a common divisor,
it is no longer guaranteed
that the order of $\cg_Y$ is a multiple of $\co a_i$.
We may, however, multiply $\cg_Y$ by any power $c_i$ of $\cg_i$
without changing the action of $\cg_Y$ on the fields in the loop. In this
way, we replace $(-b_j)$ by $(c_i\co-b_j)$ in the above formula for
the phases under $\cg_Y$. To ensure that the order of $\cg_Y$ becomes $\co U$,
we require $\gcd(c_i\co-b_j,a_i)=1$. This is the case,
for example, if $c_i$ is the product of all primes dividing $a_i$, but not
$b_j$ (in fact, $\gcd(c_i\co-b_j,t_i)=1$ would be sufficient for our purpose).
This completes the construction of the maximal phase symmetry of a skeleton
polynomial.

\subsection{All abelian symmetries of a skeleton}

Obviously the result of the procedure described above is a direct product of
cyclic groups $G=\ZZ_{n_1}\times\cdots\times\ZZ_{n_k}$.
Now we want to find all subgroups $G$ of this group that fulfil the following
criteria:\\
(1) $\det g=1$ for all $g\in G$,\\
(2) $\ZZ_d\subset G$,\\
(3) There is a non-degenerate polynomial that is invariant under $G$.\\
Since
$\ZZ_a\times\ZZ_b=\ZZ_{a\times b}$ if $\gcd(a,b)=1$ our problem reduces to
the construction of all subgroups of
$\ZZ_{p^{l_1}}\times\cdots\times\ZZ_{p^{l_k}}$, for some prime number $p$,
that fulfil these criteria.
Then, for the second condition, $\ZZ_d$ has to
be replaced by its maximal subgroup whose order is a power of $p$.
In addition, the existence of a non-degenerate polynomial with the symmetry
group $G$ has to be checked again after its subgroups corresponding to
different prime numbers have been combined.
If we denote the generators of $\ZZ_{p^{l_1}}\times\cdots\times\ZZ_{p^{l_k}}$
by $g_1,\cdots g_k$, they will have determinants
$\det g_i=\exp(2\p\,a_i/p^{m_i})$ with
$\gcd(a_i,p)=1$ and $0\le m_i\le l_i$.

We use the following algorithm for
constructing the maximal subgroup with $\det =1$: \\
(1) Find the maximal $m_i$.\\
(2) If there are $i, i'$ with $m_i=m_{i'}$ and (say) $l_i\ge l_{i'}$,
replace $g_i$ by $g_ig_{i'}^{-x}$, where $x$ is chosen in such a way that
$xa_{i'}=a_i\mod p^{m_i}$. \\
(3) Repeat this until there is only one maximal $m_i$,
then replace $g_i$ by $(g_i)^p$. \\
(4) Repeat the whole procedure until $\det g_i=1$
for all $g_i$.

Denoting the result of this construction again by
$\ZZ_{p^{l_1}}\times\cdots\times\ZZ_{p^{l_k}}$,
any subgroup will be of the form
$\ZZ_{p^{\hat l_1}}\times\cdots\times\ZZ_{p^{\hat l_{\hat k}}}$ with generators
$\hat g_i=\prod_{j=1}^kg_j^{\l_{ij}}$,
$1\le i\le \hat k$, $0\le \l_{ij}<p^{l_j}$.
Of course we severely overcount the subgroups in this way, for two reasons:
A generator $\hat g$ will be equivalent to $\hat g^\l$ if $\l$ is not
divisible by $p$, and
the set of generators $\{\hat g_a, \hat g_b\}$ is
equivalent to $\{\hat g_a \hat g_b^\l, \hat g_b\}$ for any $\l$.
The first source of overcountings can be overcome in the following way:
We note that the order of $\hat g_i$ is given by
$p^{\hat l_i}=\max_j(p^{l_j}/\gcd(\l_{ij},p^{l_j}))$. If we assume the
$j$'s to be ordered, we can denote by $\tilde j(i)$ the first $j$ for
which $p^{l_j}/\gcd(\l_{ij},p^{l_j})=p^{\hat l_i}$. This allows us to
choose the ``normalization'' $\l_{i\tilde j(i)}=p^{l_j-\hat l_i}$.
Overcountings of the second type can be avoided by demanding
$\l_{ij}<\max(1,p^{\hat l_i-\hat l_{i'}})$ if $j=\tilde j(i')$.

This implies the following algorithm for constructing each subgroup
exactly once:\\
(1) Fix the type $(\hat l_1,\cdots,\hat l_{\hat k})$ of the subgroup,
with some ordering (e.g. $\hat l_i \ge \hat l_{i+1}$). Of course
$\hat l_i \le l_i$, if the same ordering for the $l_i$'s is assumed.\\
(2) Choose $\tilde j(i)$ for each $i$, with some ordering if $l_i=l_{i+1}$
(e.g. $\tilde j(i)<\tilde j(i+1)$ if $l_i=l_{i+1}$). Thereby the
$\l_{i\tilde j(i)}$'s are determined.\\
(3) Choose all other $\l_{ij}$'s subject to
$\l_{ij}<\max(1,p^{\hat l_i-\hat l_{i'}})$ if $j=\tilde j(i')$.\\
At this point we check for the $\ZZ_d$ by explicitly  calculating all
elements of
$\ZZ_{p^{\hat l_1}}\times\cdots\times\ZZ_{p^{\hat l_{\hat k}}}$ and
comparing them with the $p$-projection of the generator of the $\ZZ_d$.
Putting together the subgroups is straightforward. The check for
non-degeneracy is based on the results of \cite{cqf} and follows the
route indicated in \cite{nms}.

\section{Results}

We have implemented these ideas in a C program, which we had running for about
6 days (real time) on a workstation to produce the results presented below.
Because of this rather reasonable computing time we did not worry about the
redundancy of calculating the same orbifolds several times for different
skeletons or even calculating equivalent moddings for a particular skeleton in
case of permutation symmetries of the potential. In general this redundancy is
not too bad, since there is an average of only 10 skeletons per configuration
and the overlap occurs mainly for groups of low orders.
It is particularly bad, however, for the $1^9$, i.e. the
potential $W=\sum_{i=1}^9 X_i^3$ --
one out of 108759 skeletons -- which alone consumed more than 80\%
of our computation time and required the calculation of
about 2 million orbifolds, producing eventually
only 23 spectra, none of which was new.
Obviously, for including all torsions one will have to work harder on this
part of the calculation.

In table I we give a detailed statistics of our results.%
\footnote{There would be no point in storing the complete information from our
calculation. We do have, however, the lists of different spectra that come
from each skeleton (the files occupy a few MB of disk space). By rerunning our
program for a particular skeleton, and with a switch set for a detailed output,
we can reconstruct the origin of any spectrum of interest. The list of spectra
can be obtained from the authors or from the data for the figures in
the LaTeX file hep-th/9211047.}
According to our
organization of the computation we list, in the first 6 columns, the results
for a fixed number of non-trivial fields, and finally combine the individual
figures. Our starting point was the list of non-degenerate configurations with
$c=9$, obtained in refs. \cite{nms,kle}.
Using the algorithms described in section 3 we have then calculated for each
configuration all inequivalent skeletons, and for each skeleton all subgroups
of the maximal phase symmetry that contain the canonical $\ZZ_d$ and
allow a non-degenerate invariant polynomial. As the $\ZZ_d$ has negative
determinant for even $N$, we have added a trivial field in that case before
restricting to determinant 1. The numbers of different skeletons and symmetries
that arise in this way are given in lines 2 and 3.
Then we list the maximal numbers of symmetries and different orbifold spectra
that can come from a single skeleton.

In the line denoted by ``spectra'' we list the total number of different
spectra obtained from all configurations with the respective number of
non-trivial fields. The majority of these spectra appear in pairs with
$\ng$ and $\na$ exchanged: the numbers of singles are given in the next line.
As the invertible skeletons play an important role in this ``mirror
symmetry'', the numbers of such skeletons and their spectra are listed
in the last two lines of table I.

\begin{figure}
\vbox{\begin{center}
\begin{tabular}{||c|cccccc|c||} \hline\hline
 fields & 4 & 5 & 6 & 7 & 8 & 9 & total      \\ \hline
 configurations & 2390 & 5165 & 2567 & 669 & 47 & 1 & 10839 \\
%degenerate ones&   14 &  418 &    3 &  17 &  0 & 0 &   452 \\
 skeletons (SK) & 7674 & 30575 & 31216 & 29257 & 7422 & 2615 & 108759 \\
% average  & 3 & 5 & 12 & 43 & 157 & 2615 & 10 \\
 symmetries    & 17833 & 53282 & 139696 & 111692 & 187641 & 2324394 & 2834538\\
orbifolds per SK &$\le$140&$\le$140&$\le$13506&$\le$2664&$\le$56632&$\le$
    2052656&$\le$2052656\\
% spectra   16502   51274   94525   79139   41731   19435    302606 with
% counted as different it they come from different skeletons ... gives an
% indication of what different orbifolds in same skeleton yield
spectra per SK &$\le$43&$\le$28&$\le$63&$\le$39&$\le$47&$\le$47&$\le$63\\
spectra (SP) & 2278 &  3182 &  2002 & 1015 & 289 & 85 & 3798\\
SP without mirror & 258 & 675 & 199 & 77 & 17 & 1 & 816\\
% no m. with N & 328 & 973  514  420  98  24  -
\hline
invertible skeletons & 4556  & 11053 & 7605 & 3406 & 564 & 115 & 27299\\
spectra from ISK & 1910 & 2259 & 1651 & 793 & 242 & 73 & 2730\\
% --aaoi.Ns      0     279     285     290      72      16        -
%ISP without IMS & 0 & 0 & 11 & 4 & 3 & 0 & 12\\
%ISP without MS  & 0 & 0 & 5 & 4 & 3 & 0 & 6\\
\hline\hline \end{tabular}\hfill\\[3mm] % \hfil
Table I: Statistics of the calculation and results. (I)SK and SP denote \\
(invertible) skeletons and spectra.  % j99: 5 1/2 days;  j9': 19h;  j8: 4h;
\end{center}}
\end{figure}

In fig. 1 we show the 800 new spectra that we found in addition to the 2998
spectra of canonical orbifolds \cite{nms,kle}.
All their Euler numbers are between  $-276$ and $480$, and the
new spectrum with the largest number of non-singlet $E_6$ representations
has $\ng=116$ and $\na=230$.  % (116,230,228)[346] (81,237,312)[318]
                  % (33,273,480)+(60,246,372)[306] (27,219,384)[246]
Note that the spectra with the largest numbers of particles have large positive
Euler number, in agreement with the  % observation
expectation that orbifolding generically increases the Euler number.

Among the new models there is a number of spectra with $p_1=p_{01}+p_{10}>0$,
namely $\ng=\na=\{3,5,9,13,21\}$ for $p_1=2$ and $\ng=\na=9$ for $p_1=6$
(we count, in this paper, spectra with different $p_1$ as different;
our plots may thus in fact show up to 5 points fewer than is
indicated in the figure captions).
In accordance with the results of section 2, all these spectra have $\chi=0$.
In contrast with the canonical orbifolds of \cite{nms}, however,
they do not all correspond to tensor products of conformal field theories, as
%The reason for this
has been discussed in section 2.
This conclusion can also be drawn
from the fact that the factors would have to be either the torus,
with all coefficients of the \PP\ equal to 1, or a model with $D=2$.
By applying our program to the 922 inequivalent skeletons of the 124
non-degenerate configurations with $D=2$ we have checked that, within our
class of orbifolds, the Hodge diamond of the K3 surface and the 2-torus are
the only possible spectra. This implies that only the last two of the above
spectra can be products and is another check for the remarkably successful
geometric interpretation of $D=2$ models \cite{g}.
Some of our spectra with $p_1>0$ have recently also been obtained
from $CP_m$ coset models with non-diagonal modular invariants \cite{aaan}.
They can be compared with the 80 non-chiral spectra we have found with
$p_1=0$, which exist for
$\ng=\na\aus\{3,7,9,10,11,\ldots,48,49,52,53,55,\ldots,179,223,251\}$.

Let us now return to the discussion of trivial fields. In addition to
compensating negative determinants, they can be used to simulate
$\ZZ_2$ torsion, i.e. $\e(g,h)=-1$, between group elements of even order.
To get an idea of what one might expect from such torsions we have added 2 (3)
trivial fields for potentials with an odd (even) number of fields,
respectively. The corresponding additional $\ZZ_2$ symmetries were
added to the set of generators of the maximal group before constructing all
subgroups with determinant 1. In this way we could generate models with several
different $\ZZ_2$ torsions without much extra effort.
For 4, 5, 6 and 7 non-trivial fields, we thus obtained 114, 393, 69 and 309
new spectra as compared with the respective cases with no additional fields
(for 8 and 9 fields the calculation was stopped because it could not
be expected to finish within a reasonable time).
The total number of spectra, however, rises only from 3798 to 3837 since there
is a large overlap of spectra for different numbers of fields.

The 39 new spectra resulting from $\ZZ_2$ torsions are shown in
fig. 2 as little circles.
They all have $\ng+\na\le66$. To give a more detailed
picture we have also included in that plot all other spectra
with $\ng+\na\le80$ by dots, the ones for orbifolds bigger than the
ones for untwisted models.
Among these 39 models we find the one with the smallest dimension of the chiral
ring.
It comes from the Fermat skeleton in $\IC_{(1,1,1,1,1,1,2,2,2)}[4]$ with 6
non-trivial and 3 trivial fields and has the spectrum $\c=b_1=0$, $\ng=\na=3$.
The symmetry that generates this spectrum is a product of 3 cyclic
groups: $g_1=\ZZ_4[0,1,0,3,2,0,0,2,0]$, $g_2=\ZZ_4[0,0,1,0,1,2,2,2,0]$ and
$g_3=\ZZ_2[0,0,0,1,0,1,1,0,1]$.
Alternatively, we can omit the trivial fields, i.e. start with the
configuration $\IC_{(1,1,1,1,1,1)}[4]$,
and twist by the group $(\ZZ_4)^2\times \ZZ_2$ with
the same action on the non-trivial fields, but now with torsions
$\e(g_1,g_2)=\e(g_2,g_3)=-1$ and $\e(g_1,g_3)=1$. Of course, we also have to
use the
appropriate signs $K_g$ and torsions with the canonical $\ZZ_4$ as discussed
in section 2.
These results give us a clear hint that realistic spectra with very low numbers
of fields can be expected in the set of models with non-trivial torsions.

To check the construction of the mirror model by Berglund and H"ubsch (BH)
\cite{bh}, which applies exactly to the invertible skeletons, we have verified
for a number of such skeletons that the inverted (or ``transposed'') skeleton
yields exactly the mirror spectra.
In addition, we have examined the mirror symmetry of the complete
set of spectra that come from invertible skeletons (let us recall
that we use the term mirror symmetry in its \naive\ sense of just rotating the
Hodge diamond; we do not check the fusion rules). The BH construction
does {\it not} imply that this space is exactly symmetric, since the mirror
of a potential containing $X^\a+XY^2$ would contain a trivial field.
Indeed, we found 12 spectra violating the mirror symmetry of
the list of spectra of invertible skeletons we produced. An explicit
check of these models shows that they are of the type described
above, and that the inverted skeleton, which contains trivial fields,
produces the correct spectrum. In fact, 6 of the missing mirror spectra are
already contained in the set of spectra from non-invertible skeletons, whereas
the other 6 are part of the above 39 spectra with $\ZZ_2$ torsions.

The situation is drastically different for non-invertible models.
Of the 1068 spectra that cannot be obtained from invertible skeletons,
810 have no mirror spectrum. Figure 3 shows the 258 remaining spectra
in this class, which do have mirrors. The ones without mirror are indicated by
small dots in fig. 3, and all of them are plotted separately in fig. 4.
It should be noted that the 258 spectra with mirrors all occur in
the ``dense'' region of all spectra in fig. 5.
It is, therefore, not unlikely
that their mirror pairings are purely accidental. A look at fig. 2 shows
that indeed in some of the low-lying regions of the plot of all spectra a high
percentage of all possible combinations of Euler numbers divisible by
4 and even $\ng+\na$ occur.
%obvious: the mirror configuration can have a number of trivial fields.
% 58 2 -112  57 3 -108  53 1 -104  56 6 -100  37 5 -64  32 12 -40
% 33 15 -36  40 22 -36  21 13 -16  6 24 36    12 42 60  4 58 108
% #spec=12   [aaoi.tot --aaoi.tot]
% 58 2 -112  57 3 -108  53 1 -104  37 5 -64  6 24 36  4 58 108
% #spec=6    [aaoi.tot --aao.tot]
% alg.tex:  1898 & cycl  ->  1911  =  3176 - 1265  (spurious: 32 0 64 counted?)
% nms.tex:                   2997  =  3176 - 179   (+1 if b01 is counted extra)
% test configs: 021436587 -> #sym=28 (not 418); 012222222 about 50 min on SUN.
%                     #skel  #sym     sym/skel  #spec   spec/skel
% j90={91...96}       2400   114310   3693      16914     28
% j97                  214   157428   56632      2498     47
% j99                    1  2052656     -          23      -

Finally, we come to the presentation of what seem to be the most interesting
models.
In table II we list the 25 new spectra with a net number of 3 generations,
together with a representation as a LG orbifold.
The 40 untwisted 3-generation models were already given in
refs.~\cite{nms,kle}:
they have 29 different spectra with
$\ng=\na-3\aus\{13,15,16,17,20,26,27,29,31,34,37,\lbo42,\lbo47,\lbo67,\lbo74\}$
and $\na=\ng-3\aus\{13,17,20,21,23,26,29,32,35,37,40,\lbo47,\lbo48,\lbo57\}$
for positive and
negative Euler number, respectively.
In view of the much smaller increase of the total number of spectra, this
shows again that the percentage of ``realistic'' models is larger for
%become much more abundant with ("werden viel reichlicher")
more general constructions.

\begin{figure}
\vbox{\noindent
\let\Atop=\atop % \def\[{\left[\displaystyle} \def\Atop{\atop\displaystyle }
\begin{center} \let\komma=\, \def\,{\komma\komma}
\begin{tabular}{||c@{\V~~}cr|l|l||} \hline\hline
$\ng$&$\na$&$\chi$& configuration & twist \\ \hline
 14 & 11 &$-6$& $\IC_{(2,3,3,4,4,5,5)}[13]$ & $\ZZ_2[1\,1\,0\,1\,0\,0\,1]$ \\
 17 & 14 &$-6$& $\IC_{(2,2,2,3,3,3,3)}[9]$ &
    $(\ZZ_3)^2\[0\,0\,1\,0\,0\,1\,1\Atop0\,1\,0\,0\,0\,2\,0\]$\\
 18 & 15 &$-6$& $\IC_{(1,2,3,3,3,3,3)}[9]$ & $\ZZ_4[1\,0\,2\,0\,3\,0\,0]$\\
 21 & 18 &$-6$& $\IC^I_{(2,3,5,8,9)}[27]$ & $\ZZ_2[0\,1\,0\,1\,0]$\\
 25 & 22 &$-6$& $\IC_{(2,3,4,9,9)}[27]$ & $\ZZ_2[0\,0\,1\,0\,1]$\\
 30 & 27 &$-6$& $\IC^I_{(1,4,5,5,10)}[25]$ & $\ZZ_2[1\,0\,0\,0\,1]$ \\
 31 & 28 &$-6$& $\IC_{(1,3,4,4,9)}[21]$ & $\ZZ_3[0\,0\,1\,2\,0]$ \\
 34 & 31 &$-6$& $\IC^I_{(1,2,3,3,8)}[17]$ &
		$(\ZZ_2)^2\[0\,0\,0\,1\,1\Atop0\,1\,1\,0\,0\]$ \\
 37 & 34 &$-6$& $\IC_{(2,3,5,15,20)}[45]$ & $\ZZ_2\[0\,1\,0\,0\,1\]$ \\
 45 & 42 &$-6$& $\IC^I_{(1,3,9,14,18)}[45]$ & $\ZZ_2\[1\,0\,0\,0\,1\]$ \\
 47 & 44 &$-6$& $\IC_{(2,3,7,21,30)}[63]$ & $\ZZ_2\[1\,0\,1\,0\,0\]$ \\
 49 & 46 &$-6$& $\IC^I_{(1,2,5,10,17)}[35]$ & $\ZZ_3\[0\,2\,0\,1\,0\]$ \\
 54 & 51 &$-6$& $\IC_{(1,2,4,13,19)}[39]$ & $\ZZ_4\[2\,2\,3\,0\,1\]$ \\
 66 & 63 &$-6$& $\IC_{(1,3,11,18,32)}[65]$ & $\ZZ_2\[0\,1\,1\,1\,1\]$ \\
 70 & 67 &$-6$& $\IC^I_{(1,3,15,20,36)}[75]$ & $\ZZ_2\[1\,0\,0\,0\,1\]$ \\
 14 & 17 & 6 &$\IC_{(1,1,1,1,2,2,2)}[5]$&$\ZZ_{16}[8\,12\,6\,13\,14\,10\,1]$\\
 18 & 21 & 6 & $\IC^I_{(2,3,5,8,9)}[27]$ & $\ZZ_2\[1\,0\,1\,0\,0\]$ \\
 19 & 22 & 6 & $\IC_{(2,3,3,5,5,6,6)}[15]$ & $\ZZ_2\[0\,0\,1\,0\,0\,1\,0\]$ \\
 21 & 24 & 6 & $\IC^I_{(1,1,2,2,5)}[11]$ & $\ZZ_5\[0\,1\,1\,1\,2\]$ \\
 23 & 26 & 6 & $\IC_{(1,2,4,5,7)}[19]$ & $\ZZ_3\[1\,2\,0\,0\,0\]$ \\
 32 & 35 & 6 & $\IC^I_{(3,3,10,14,15)}[45]$ & $\ZZ_2\[0\,1\,0\,1\,0\]$ \\
 36 & 39 & 6 & $\IC_{(1,2,3,9,12)}[27]$ &
		      $(\ZZ_2)^2\[1\,1\,0\,0\,0\Atop1\,0\,0\,0\,1\]$ \\
 40 & 43 & 6 & $\IC^I_{(1,1,5,8,10)}[25]$ &
		      $(\ZZ_2)^2\[1\,0\,0\,1\,0\Atop0\,1\,0\,0\,1\]$ \\
 46 & 49 & 6 & $\IC^I_{(1,6,7,21,28)}[63]$ & $\ZZ_2\[1\,0\,0\,0\,1\]$ \\
 48 & 51 & 6 & $\IC^I_{(1,5,6,18,25)}[55]$ & $\ZZ_2\[1\,0\,1\,1\,1\]$ \\
\hline\hline \end{tabular}\hfill\\[3mm] \hfil
Table II: New 3-generation models. A superscript $I$ indicates that the twist
can be applied\\
	  at an invertible point in the moduli space of the configuration.
\end{center}}\end{figure}

The first model in table~II is the one with the lowest $\ng+\na$.
It belongs to the configuration $\IC_{(2,3,3,4,4,5,5)}[13]$ and is
represented by the non-invertible potential
\beq W=X^5Y+Y^3U+Z^3V+U^2W+V^2W+W^2Z+T^2Z+\e_1UVT+\e_2WTY+\e_3WX^4. \eql{3g}
%skeleton $1345522$ or $534511$; configuration $\IC_{(2,3,3,4,4,5,5)}[13]$,
%First skeleton: $(UV)[W]=UVT$, $(WT)[Z]=WTY$ $\then$ $(WTX)[YZ]=WX^4$.
The twist by $\ZZ_2$[1 1 0 1 0 0 1] with $X_i=\{X,Y,Z,U,V,W,T\}$ transforms
its spectrum from $(29,5,-48)$ into $(14,11,-6)$. This indicates that one
should be careful in applying empirical ``quantization rules'' in the
search for 3-generation models.
Actually, this model corresponds to two different skeletons, both of which
belong to the same non-degenerate potential
(in the language of \cite{cqf}, each of the monomials $X^5Y$ and
$X^4W$ in (\ref{3g}) can be interpreted as a pointer; then the other one
belongs to the set of required links).

In table III we list the 1-generation models. None of them has a mirror
spectrum, and hence none of them comes from an invertible skeleton. With two
exceptions, all values of the original inverse charge quantum $d$ are prime.
In fact, as for the untwisted case~\cite{nms}, all 1- and 3-generation models
have odd $d$, and hence an odd number of fields.
The number of 2-generation models in our list is 33; 26 of them do not
require a twist and 18 have negative Euler numbers.

\begin{figure}
\vbox{\noindent\begin{center}
\begin{tabular}{||c@{\V~~}cr|l|l||} \hline\hline
$\ng$&$\na$&$\chi$& configuration & twist \\ \hline
 12  & 13  & 2 & $\IC_{(5,6,7,8,9,11,12)}[29]$ & untwisted\\
 16  & 17  & 2 & $\IC_{(3,3,4,5,7,8,8)}[19]$ & $\ZZ_2$[0 1 0 0 0 1 0]\\
 20  & 21  & 2 & $\IC_{(3,3,4,5,10)}[25]$ & $\ZZ_2$[0 1 0 0 1]\\
 22  & 23  & 2 & $\IC_{(3,4,5,7,16)}[35]$ & $\ZZ_2$[0 1 0 0 1]\\
 34  & 35  & 2 & $\IC_{(1,1,1,4,6)}[13]$ & $\ZZ_6$[0 2 1 4 5]\\
 22  & 21  &$-2$& $\IC_{(2,3,4,5,9)}[23]$ & $\ZZ_2$[1 0 0 0 1]\\
 33  & 32  &$-2$& $\IC_{(1,3,4,4,11)}[23]$ & $\ZZ_3$[0 0 2 1 0]\\
 46  & 45  &$-2$& $\IC_{(1,5,6,12,23)}[47]$ & $\ZZ_2$[0 1 0 1 0]\\
\hline\hline \end{tabular}\hfill\\[3mm]
Table III: 1-generation models
\end{center}} \end{figure}

\del Large $\chi$ -- no mirror (all untwisted?)
\begin{center}
\begin{tabular}{||c|ccccccccccccccc||} \hline\hline
$\ng$&13&17&20&7&14&27&36&15&38&16&36&37&39& 278&\ldots\\
$\na$&433&341&326&295&284&291&294&271&293&254&270&265&264& 53&\ldots\\
$\chi$&840&648&612&576&540&528&516&512&510&476&468&456&450& $-450$&\ldots\\
\hline\hline \end{tabular}\hfill\\[3mm]
of $\chi$.
\end{center}
% 8 206 396   39 264 450  37 265 456   36 270 468   16 254 476   38 293 510
% 15 271 512  36 294 516  27 291 528   14 284 540   7 295 576    20 326 612
% 17 341 648  13 433 840  278 53 -450  276 54 -444  235 27 -416  249 54 -390
\enddel

It is significant that the spectra without mirror that have the largest values
of $|\chi|$ all come from untwisted LG models (compare figs.~1 and 4).
The first 12 of these,
with spectra $(13,433)$, $(17,341)$, (20,326), $\ldots$, have large positive
Euler numbers. As all our new spectra have much smaller particle content,
it appears to be most unlikely that these spectra will eventually find their
mirrors in the realm of (more general) LG orbifolds.

\section{Discussion and outlook}

Considering the complete set of abelian symmetries of Landau--Ginzburg
potentials, we have studied approximately 250 times as many models as
previously, with canonically twisted theories
(there is, however, some redundancy in our constructions owing to the
possible occurrence of the same symmetries for different skeletons in a
specific configuration and to permutation symmetries of some skeletons).
Doing so
without pushing computer time to astronomical heights was only possible
with an algorithm that was extremely efficient, at least at its central part,
i.e. at the calculation of the numbers of chiral generations and
anti-generations from a given potential and symmetry.
The numbers of spectra
obtained, and the number of new features, however, did not rise in a
comparable manner. Less than 25\% of the spectra we found were new
compared with \cite{nms,kle},
and the overall impression of the plot of spectra in fig.~5 is
the same as in the pioneering work ref.~\cite{cls}.
Although the new spectra arise primarily in the range of
low generation and anti-generation numbers, we have not found any
models that look particularly promising from a phenomenological point
of view. Yet, we believe that
our results are quite interesting from the following points of view:

We have established now what we already found in \cite{nms},
namely that mirror symmetry in the context of Landau--Ginzburg orbifolds
occurs regularly for those and only for those models for which the
Berglund--H\"ubsch construction works.
In particular, among the spectra that remain without mirrors, there are a
number of canonical LG models whose Euler numbers are
much larger than what any orbifold contributed.
This makes it appear very unlikely
that a generalization of the BH contruction exists for LG orbifolds or
for the related Calabi--Yau manifolds.
In accordance with the results of~\cite{fk}, the lack of mirror symmetry also
indicates that, for a complete classification of rational $N=2$ theories,
we need to go beyond LG models.  Still, in addition to their phenomenological
use, the lessons we learn from them may be helpful also in that direction.

Obviously non-abelian symmetries and twists with non-trivial
torsion~\cite{fiqs} are good candidates for providing phenomenologically more
realistic spectra. This is clear from the fact that our ``smallest''
3-generation model requires nearly twice as many
fermions of opposite chirality
as the well-known model of \cite{rolf}, which
can be interpreted as a non-abelian Landau--Ginzburg orbifold.
Furthermore, the few new spectra that we obtained with our
excursion into the space of models with non-trivial $\ZZ_2$ torsion
are all in the area of very small particle numbers, and this set contains
the model with the least value of $\ng+\na$ that we have found.

There are severe obstacles to a complete classification of models
with torsion along the lines of this work: Our fast algorithm for the
calculation of spectra cannot be generalized to the case of torsion.
It relies on the fact that, without torsion, the only characteristic
of a group element required for the calculation of the spectrum is
the information about which subset of the fields it leaves invariant.
Besides, the number of possible different torsions for a set of $n$
generators of order $p$ is given by $p^{({n\atop 2})}$. For the $1^9$
with the maximal symmetry  we have, with unit determinants
and the restrictions on torsions with $\ZZ_d$,
$p=3$ and $n=7$, thus
yielding $3^{21}$ cases, which makes a \naive\
calculation completely impossible.
Still, our analysis and computation of abelian symmetries provides the
necessary first step for such an investigation, be it complete or not.

With our present knowledge, it seems that the number of consistent spectra
of $N=2$ superconformal field theories with $c=9$ and integer charges
is restricted to a few
thousands. This raises the question of whether it might not be possible to
classify all (2,2) vacua by enumeration, with a scheme based only
on the axioms of $N=2$ superconformal field theory. The mathematical
tools for such a task, however, are still waiting to be discovered.

{\it Acknowledgements.} It is a pleasure to thank Per Berglund and Philip
Candelas for discussions and Tristan H"ubsch for correspondence.

\newpage \noindent {\Large\bf Figures} \bsk\bsk

% ========================     Fig. 1     ========================== %
\def\figsca{\funit=0.333truemm \wi=500 \he=350}
\def\figlab{\hlab{-200} \hlab{200} \hlab{-400} \hlab{400}
            \hlab{-100} \hlab{100} \hlab{-300} \hlab{300}
	    \vlab{100}  \vlab{200} \vlab{300}  }
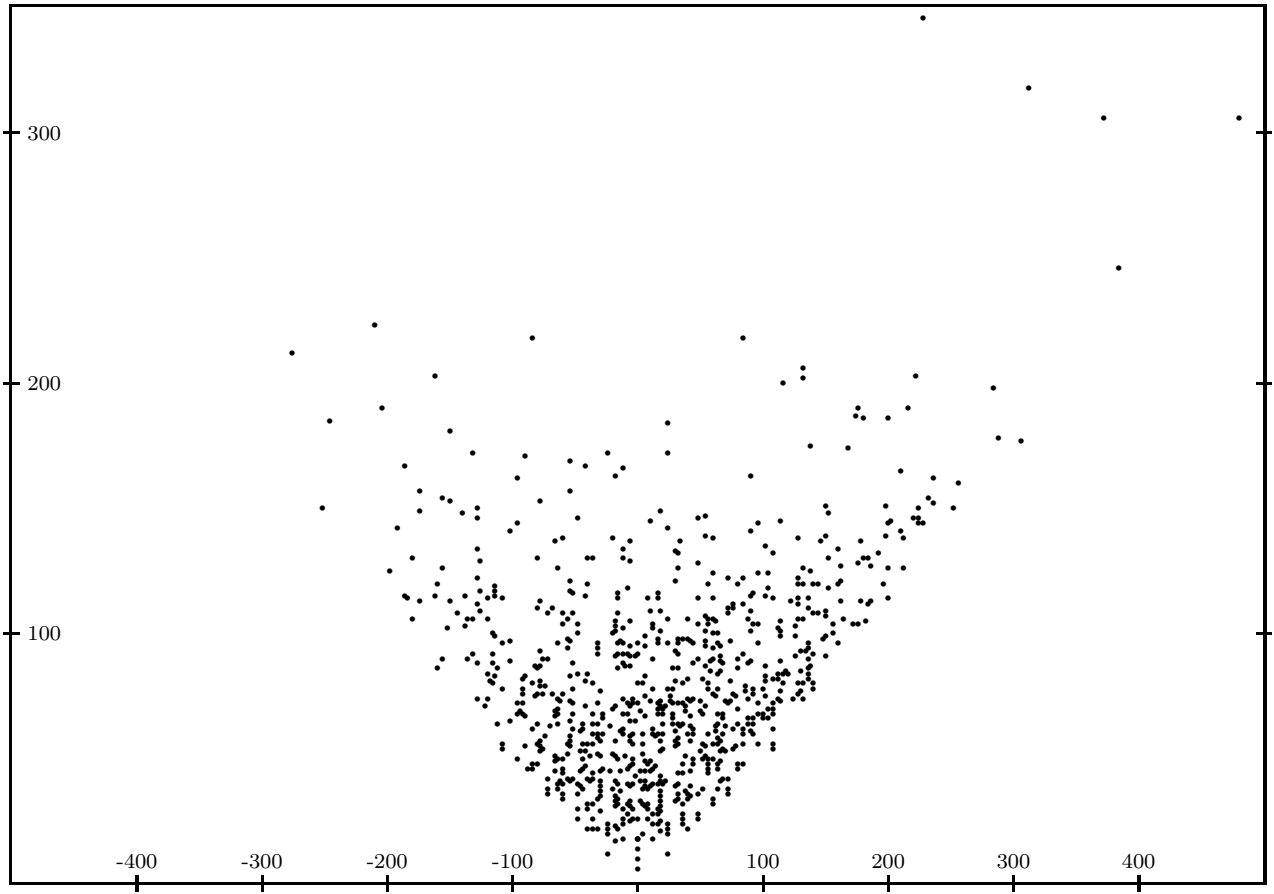
\def\figcap{$\ng+\na$ vs. Euler number $\chi$ for the 800 new LGO spectra}
\Vplo{
\.175,37)\.138,12)\.154,31)\.164,59)\.146,44)\.112,13)\.119,23)\.104,11)
\.130,37)\.103,11)\.98,8)\.110,20)\.100,13)\.118,31)\.122,35)\.98,17)\.142,61)
\.83,3)\.100,20)\.84,6)\.102,24)\.116,38)\.89,13)\.94,19)\.114,39)\.128,53)
\.90,18)\.109,39)\.86,17)\.92,23)\.79,11)\.87,19)\.79,13)\.86,20)\.119,53)
\.69,5)\.76,12)\.88,24)\.93,29)\.99,35)\.105,41)\.107,43)\.86,23)\.90,27)
\.96,33)\.66,5)\.67,7)\.72,12)\.83,23)\.87,27)\.70,11)\.69,11)\.73,15)\.75,17)
\.79,21)\.70,13)\.78,21)\.86,29)\.87,30)\.88,31)\.60,4)\.71,15)\.54,0)\.55,1)
\.66,12)\.75,21)\.84,30)\.58,7)\.70,19)\.74,23)\.96,45)\.49,1)\.58,10)\.60,12)
\.96,48)\.105,57)\.58,11)\.57,11)\.59,13)\.61,15)\.62,16)\.64,18)\.50,5)
\.56,11)\.64,19)\.108,63)\.45,1)\.44,2)\.45,3)\.52,10)\.61,19)\.130,88)
\.58,17)\.64,23)\.44,4)\.48,8)\.52,12)\.58,18)\.63,23)\.75,35)\.85,45)\.46,7)
\.47,8)\.48,9)\.59,20)\.60,21)\.63,24)\.66,27)\.76,37)\.96,57)\.46,8)\.57,19)
\.64,26)\.48,11)\.58,21)\.36,0)\.37,1)\.39,3)\.63,27)\.72,36)\.49,14)\.55,21)
\.72,38)\.39,6)\.41,8)\.42,9)\.50,17)\.51,18)\.85,52)\.35,3)\.36,4)\.41,9)
\.48,16)\.50,18)\.51,19)\.53,21)\.64,32)\.79,47)\.36,5)\.52,21)\.32,2)\.33,3)
\.35,5)\.37,7)\.38,8)\.40,10)\.53,23)\.67,37)\.69,39)\.84,54)\.35,7)\.40,12)
\.42,14)\.61,33)\.63,35)\.67,39)\.34,7)\.41,14)\.42,15)\.43,16)\.46,19)
\.50,23)\.55,28)\.62,35)\.72,45)\.74,47)\.92,65)\.98,71)\.34,8)\.44,18)
\.47,21)\.49,23)\.52,26)\.57,31)\.67,41)\.71,45)\.25,1)\.27,3)\.30,6)\.32,8)
\.54,30)\.62,38)\.64,40)\.85,61)\.31,8)\.34,11)\.38,15)\.42,19)\.30,8)\.34,12)
\.36,14)\.39,17)\.42,20)\.43,21)\.34,13)\.37,16)\.46,25)\.51,30)\.68,47)
\.94,73)\.21,1)\.25,5)\.26,6)\.31,11)\.38,18)\.52,32)\.70,50)\.75,55)\.30,11)
\.35,16)\.20,2)\.22,4)\.25,7)\.30,12)\.31,13)\.37,19)\.39,21)\.41,23)\.43,25)
\.49,31)\.74,56)\.19,3)\.25,9)\.34,18)\.37,21)\.38,22)\.40,24)\.44,28)\.54,38)
\.55,39)\.56,40)\.22,7)\.25,10)\.26,11)\.27,12)\.28,13)\.30,15)\.33,18)
\.36,21)\.46,31)\.30,16)\.37,23)\.40,26)\.41,27)\.12,0)\.16,4)\.17,5)\.18,6)
\.92,80)\.28,17)\.37,26)\.24,14)\.31,21)\.40,30)\.55,45)\.74,64)\.13,4)
\.16,7)\.20,11)\.21,12)\.22,13)\.25,16)\.30,21)\.33,24)\.40,31)\.50,41)
\.55,46)\.56,47)\.57,48)\.86,77)\.15,7)\.20,12)\.21,13)\.24,16)\.29,21)
\.47,39)\.50,42)\.52,44)\.58,50)\.61,53)\.62,54)\.22,15)\.26,19)\.34,27)
\.52,45)\.12,6)\.15,9)\.16,10)\.18,12)\.33,27)\.34,28)\.37,31)\.40,34)\.47,41)
\.49,43)\.51,45)\.54,48)\.68,62)\.70,64)\.86,80)\.28,23)\.46,41)\.16,12)
\.18,14)\.22,18)\.25,21)\.36,32)\.38,34)\.48,44)\.61,57)\.14,11)\.17,14)
\.18,15)\.21,18)\.25,22)\.30,27)\.31,28)\.34,31)\.37,34)\.45,42)\.47,44)
\.49,46)\.54,51)\.66,63)\.70,67)\.14,12)\.16,14)\.21,19)\.25,23)\.26,24)
\.29,27)\.31,29)\.37,35)\.38,36)\.22,21)\.33,32)\.46,45)\.7,7)\.9,9)\.36,36)
\.40,40)\.46,46)\.48,48)\.16,17)\.20,21)\.22,23)\.34,35)\.9,11)\.15,17)
\.18,20)\.25,27)\.27,29)\.29,31)\.39,41)\.14,17)\.18,21)\.19,22)\.21,24)
\.23,26)\.32,35)\.36,39)\.40,43)\.46,49)\.48,51)\.11,15)\.13,17)\.14,18)
\.17,21)\.22,26)\.55,59)\.17,22)\.20,25)\.22,27)\.34,39)\.52,57)\.70,75)
\.6,12)\.9,15)\.11,17)\.17,23)\.20,26)\.27,33)\.28,34)\.36,42)\.42,48)\.48,54)
\.49,55)\.20,27)\.26,33)\.8,16)\.10,18)\.11,19)\.16,24)\.26,34)\.30,38)
\.31,39)\.32,40)\.44,52)\.45,53)\.53,61)\.54,62)\.6,15)\.8,17)\.10,19)\.11,20)
\.12,21)\.13,22)\.14,23)\.16,25)\.17,26)\.22,31)\.24,33)\.31,40)\.32,41)
\.46,55)\.50,59)\.70,79)\.15,25)\.22,32)\.25,35)\.27,37)\.29,39)\.30,40)
\.15,26)\.30,41)\.0,12)\.4,16)\.5,17)\.6,18)\.33,45)\.42,54)\.47,59)\.65,77)
\.80,92)\.86,98)\.30,43)\.31,44)\.25,39)\.26,40)\.27,41)\.29,43)\.32,46)
\.36,50)\.9,24)\.12,27)\.20,35)\.21,36)\.23,38)\.33,48)\.39,54)\.53,68)
\.59,74)\.9,25)\.10,26)\.12,28)\.14,30)\.20,36)\.21,37)\.24,40)\.28,44)
\.35,51)\.38,54)\.40,56)\.41,57)\.55,71)\.58,74)\.23,40)\.60,77)\.3,21)
\.4,22)\.5,23)\.7,25)\.13,31)\.15,33)\.23,41)\.27,45)\.31,49)\.40,58)\.9,28)
\.16,35)\.25,44)\.26,45)\.1,21)\.7,27)\.8,28)\.20,40)\.27,47)\.31,51)\.39,59)
\.43,63)\.7,28)\.9,30)\.12,33)\.18,39)\.21,42)\.23,44)\.26,47)\.38,59)\.13,35)
\.14,36)\.19,41)\.20,42)\.26,48)\.33,55)\.37,59)\.1,25)\.2,26)\.6,30)\.33,57)
\.40,64)\.45,69)\.52,76)\.61,85)\.14,39)\.22,47)\.24,49)\.6,32)\.12,38)
\.15,41)\.21,47)\.28,54)\.12,39)\.14,41)\.22,49)\.30,57)\.34,61)\.35,62)
\.37,64)\.40,67)\.56,83)\.60,87)\.8,36)\.9,37)\.10,38)\.11,39)\.16,44)\.25,53)
\.26,54)\.38,66)\.39,67)\.46,74)\.28,57)\.30,59)\.1,31)\.2,32)\.10,40)\.13,43)
\.23,53)\.30,60)\.32,62)\.35,65)\.38,68)\.42,72)\.47,77)\.54,84)\.15,46)
\.16,47)\.22,53)\.37,68)\.3,35)\.7,39)\.8,40)\.12,44)\.13,45)\.16,48)\.18,50)
\.23,55)\.26,58)\.32,64)\.34,66)\.4,37)\.8,41)\.10,43)\.12,45)\.26,59)\.28,61)
\.29,62)\.31,64)\.4,38)\.10,44)\.13,47)\.16,50)\.17,51)\.27,61)\.9,44)\.14,49)
\.19,54)\.0,36)\.1,37)\.3,39)\.18,54)\.36,72)\.37,73)\.43,79)\.18,55)\.22,59)
\.30,67)\.8,46)\.12,50)\.19,57)\.36,74)\.37,75)\.8,47)\.18,57)\.3,43)\.4,44)
\.12,52)\.15,55)\.23,63)\.25,65)\.40,80)\.3,45)\.7,49)\.9,51)\.10,52)\.25,67)
\.35,77)\.40,82)\.88,130)\.17,60)\.18,61)\.10,54)\.11,55)\.15,59)\.22,66)
\.31,75)\.8,53)\.22,67)\.28,73)\.32,77)\.35,80)\.48,93)\.59,104)\.7,53)
\.9,55)\.10,56)\.15,61)\.16,62)\.29,75)\.35,81)\.4,52)\.10,58)\.12,60)\.24,72)
\.28,76)\.38,86)\.48,96)\.8,58)\.9,59)\.14,64)\.12,63)\.15,66)\.16,67)\.18,69)
\.32,83)\.42,93)\.7,59)\.9,61)\.33,85)\.36,88)\.0,54)\.1,55)\.4,58)\.7,61)
\.8,62)\.9,63)\.14,68)\.30,84)\.39,93)\.9,65)\.13,69)\.14,70)\.23,79)\.8,65)
\.10,67)\.16,73)\.21,78)\.22,79)\.24,81)\.44,101)\.11,69)\.13,71)\.71,129)
\.13,72)\.12,72)\.26,87)\.6,68)\.14,77)\.18,81)\.20,83)\.6,70)\.8,72)\.21,85)
\.24,88)\.25,89)\.28,92)\.29,93)\.37,101)\.6,71)\.10,75)\.14,79)\.4,70)
\.7,73)\.20,86)\.27,93)\.30,96)\.68,134)\.70,136)\.13,80)\.7,75)\.8,76)
\.9,77)\.11,79)\.13,81)\.14,82)\.21,89)\.23,91)\.9,78)\.28,97)\.53,122)
\.4,74)\.5,75)\.11,81)\.19,89)\.25,95)\.18,90)\.24,96)\.32,105)\.12,86)
\.8,83)\.12,87)\.16,91)\.17,92)\.32,107)\.38,113)\.21,97)\.27,103)\.36,112)
\.11,89)\.13,91)\.8,88)\.20,100)\.27,107)\.16,97)\.20,101)\.23,104)\.12,94)
\.45,129)\.9,95)\.50,137)\.8,96)\.20,108)\.51,139)\.12,101)\.24,113)\.20,110)
\.48,138)\.7,98)\.10,102)\.19,111)\.10,103)\.17,110)\.18,114)\.11,109)\.20,119)
\.26,125)\.7,107)\.13,113)\.22,122)\.43,143)\.22,123)\.18,123)\.30,135)
\.10,116)\.16,122)\.41,149)\.18,128)\.46,157)\.16,128)\.17,129)\.19,131)
\.15,129)\.116,230)\.19,135)\.17,135)\.22,140)\.12,138)\.16,144)\.28,170)
\.17,161)\.12,165)\.81,237)\.60,246)\.27,219)\.33,273)\.3,3)\.5,5)\.9,9)
\.13,13)\.9,9) % #spec=800
}
\newpage

% ========================     Fig. 2     ========================== %
\def\figsca{\funit=1truemm \wi=165 \he=80}
\def\figlab{\hlab{-50} \hlab{50} \hlab{-100} \hlab{100} \hlab{-150} \hlab{150}
	    \vlab{20}  \vlab{40} \vlab{60} }
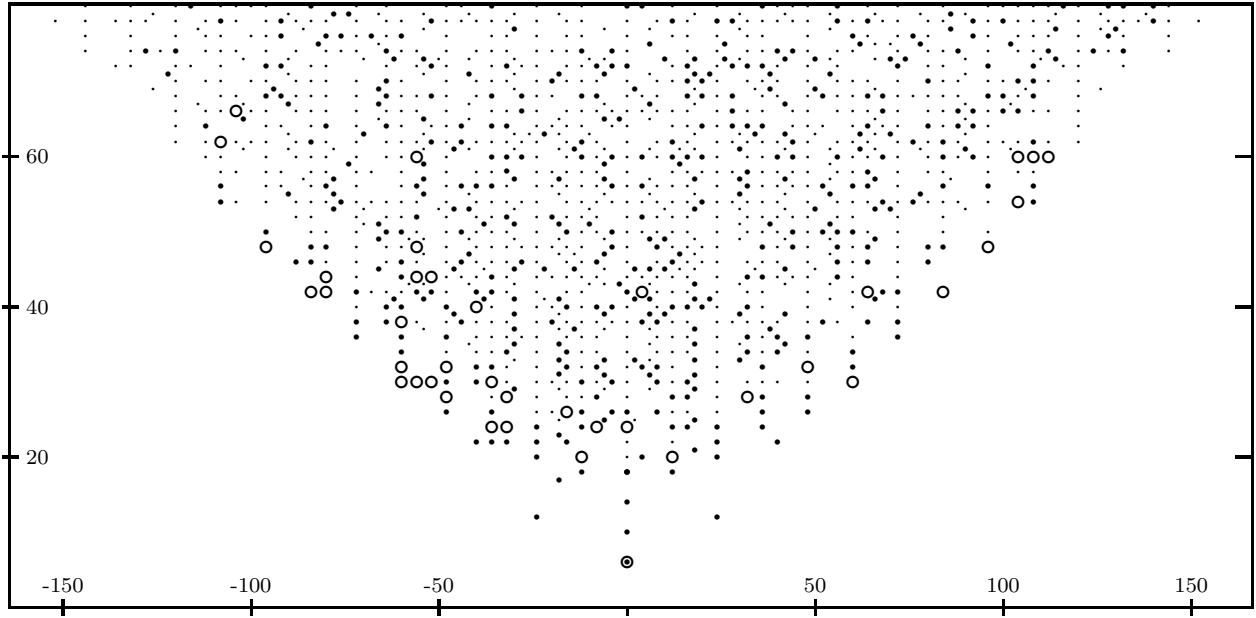
\def\figcap{$\ng+\na$ vs. Euler number for the 39 spectra with
	    $\ZZ_2$ torsion (circles)\\~~~~~~and for the LG/LGO spectra with
	    $\ng+\na\le80$ (small/big dots)}
\Vplo{
\def\npt{\funit=1pt\circle4}  % new LGO points with trivial (i.e. Z_2 torsion)
\.58,4)\.59,7)\.48,0)\.42,0)\.41,1)\.42,2)\.30,0)\.31,1)\.34,4)\.29,1)\.36,8)
\.38,10)\.44,16)\.28,2)\.35,9)\.26,2)\.28,4)\.30,10)\.21,3)\.24,6)\.20,4)
\.22,6)\.17,9)\.13,7)\.14,10)\.3,3)\.12,12)\.20,22)\.7,13)\.6,22)\.4,28)
\.0,30)\.5,37)\.0,42)\.0,48)\.1,53)\.4,56)\.3,57)\.2,58) % #spec=39
\def\npt{\funit=1pt\circle*2} % now LGO points without trivial && g+a < 81
\.69,5)\.66,5)\.67,7)\.69,11)\.60,4)\.54,0)\.55,1)\.66,12)\.58,7)\.49,1)
\.58,10)\.60,12)\.58,11)\.57,11)\.59,13)\.61,15)\.62,16)\.50,5)\.56,11)
\.45,1)\.44,2)\.45,3)\.52,10)\.61,19)\.58,17)\.44,4)\.48,8)\.52,12)\.58,18)
\.46,7)\.47,8)\.48,9)\.59,20)\.46,8)\.57,19)\.48,11)\.58,21)\.36,0)\.37,1)
\.39,3)\.49,14)\.55,21)\.39,6)\.41,8)\.42,9)\.50,17)\.51,18)\.35,3)\.36,4)
\.41,9)\.48,16)\.50,18)\.51,19)\.53,21)\.36,5)\.52,21)\.32,2)\.33,3)\.35,5)
\.37,7)\.38,8)\.40,10)\.53,23)\.35,7)\.40,12)\.42,14)\.34,7)\.41,14)\.42,15)
\.43,16)\.46,19)\.50,23)\.34,8)\.44,18)\.47,21)\.49,23)\.52,26)\.25,1)\.27,3)
\.30,6)\.32,8)\.31,8)\.34,11)\.38,15)\.42,19)\.30,8)\.34,12)\.36,14)\.39,17)
\.42,20)\.43,21)\.34,13)\.37,16)\.46,25)\.21,1)\.25,5)\.26,6)\.31,11)\.38,18)
\.30,11)\.35,16)\.20,2)\.22,4)\.25,7)\.30,12)\.31,13)\.37,19)\.39,21)\.41,23)
\.43,25)\.49,31)\.19,3)\.25,9)\.34,18)\.37,21)\.38,22)\.40,24)\.44,28)\.22,7)
\.25,10)\.26,11)\.27,12)\.28,13)\.30,15)\.33,18)\.36,21)\.46,31)\.30,16)
\.37,23)\.40,26)\.41,27)\.12,0)\.16,4)\.17,5)\.18,6)\.28,17)\.37,26)\.24,14)
\.31,21)\.40,30)\.13,4)\.16,7)\.20,11)\.21,12)\.22,13)\.25,16)\.30,21)\.33,24)
\.40,31)\.15,7)\.20,12)\.21,13)\.24,16)\.29,21)\.22,15)\.26,19)\.34,27)
\.12,6)\.15,9)\.16,10)\.18,12)\.33,27)\.34,28)\.37,31)\.40,34)\.28,23)\.16,12)
\.18,14)\.22,18)\.25,21)\.36,32)\.38,34)\.14,11)\.17,14)\.18,15)\.21,18)
\.25,22)\.30,27)\.31,28)\.34,31)\.37,34)\.14,12)\.16,14)\.21,19)\.25,23)
\.26,24)\.29,27)\.31,29)\.37,35)\.38,36)\.22,21)\.33,32)\.7,7)\.9,9)\.36,36)
\.40,40)\.16,17)\.20,21)\.22,23)\.34,35)\.9,11)\.15,17)\.18,20)\.25,27)
\.27,29)\.29,31)\.39,41)\.14,17)\.18,21)\.19,22)\.21,24)\.23,26)\.32,35)
\.36,39)\.11,15)\.13,17)\.14,18)\.17,21)\.22,26)\.17,22)\.20,25)\.22,27)
\.34,39)\.6,12)\.9,15)\.11,17)\.17,23)\.20,26)\.27,33)\.28,34)\.36,42)\.20,27)
\.26,33)\.8,16)\.10,18)\.11,19)\.16,24)\.26,34)\.30,38)\.31,39)\.32,40)
\.6,15)\.8,17)\.10,19)\.11,20)\.12,21)\.13,22)\.14,23)\.16,25)\.17,26)\.22,31)
\.24,33)\.31,40)\.32,41)\.15,25)\.22,32)\.25,35)\.27,37)\.29,39)\.30,40)
\.15,26)\.30,41)\.0,12)\.4,16)\.5,17)\.6,18)\.33,45)\.30,43)\.31,44)\.25,39)
\.26,40)\.27,41)\.29,43)\.32,46)\.9,24)\.12,27)\.20,35)\.21,36)\.23,38)
\.9,25)\.10,26)\.12,28)\.14,30)\.20,36)\.21,37)\.24,40)\.28,44)\.23,40)
\.3,21)\.4,22)\.5,23)\.7,25)\.13,31)\.15,33)\.23,41)\.27,45)\.31,49)\.9,28)
\.16,35)\.25,44)\.26,45)\.1,21)\.7,27)\.8,28)\.20,40)\.27,47)\.7,28)\.9,30)
\.12,33)\.18,39)\.21,42)\.23,44)\.26,47)\.13,35)\.14,36)\.19,41)\.20,42)
\.26,48)\.1,25)\.2,26)\.6,30)\.14,39)\.22,47)\.24,49)\.6,32)\.12,38)\.15,41)
\.21,47)\.12,39)\.14,41)\.22,49)\.8,36)\.9,37)\.10,38)\.11,39)\.16,44)\.25,53)
\.26,54)\.1,31)\.2,32)\.10,40)\.13,43)\.23,53)\.15,46)\.16,47)\.22,53)\.3,35)
\.7,39)\.8,40)\.12,44)\.13,45)\.16,48)\.18,50)\.23,55)\.4,37)\.8,41)\.10,43)
\.12,45)\.4,38)\.10,44)\.13,47)\.16,50)\.17,51)\.9,44)\.14,49)\.19,54)\.0,36)
\.1,37)\.3,39)\.18,54)\.18,55)\.8,46)\.12,50)\.19,57)\.8,47)\.18,57)\.3,43)
\.4,44)\.12,52)\.15,55)\.3,45)\.7,49)\.9,51)\.10,52)\.17,60)\.18,61)\.10,54)
\.11,55)\.15,59)\.8,53)\.7,53)\.9,55)\.10,56)\.15,61)\.16,62)\.4,52)\.10,58)
\.12,60)\.8,58)\.9,59)\.14,64)\.12,63)\.7,59)\.9,61)\.0,54)\.1,55)\.4,58)
\.7,61)\.8,62)\.9,63)\.9,65)\.8,65)\.10,67)\.11,69)\.6,68)\.6,70)\.8,72)
\.6,71)\.4,70)\.7,73)\.4,74)\.5,75)\.3,3)\.5,5)\.9,9)\.13,13)\.21,21)\.9,9)
% #spec=428
\def\npt{{\funit=1pt \circle*1}} % now LG points with g+a<81
\.77,1)\.73,1)\.74,2)\.75,3)\.76,4)\.70,2)\.73,5)\.69,3)\.70,4)\.71,5)\.72,6)
\.73,7)\.71,7)\.66,3)\.71,8)\.68,6)\.61,1)\.62,2)\.63,3)\.64,4)\.65,5)\.66,6)
\.68,8)\.69,9)\.70,10)\.65,7)\.62,5)\.67,10)\.58,2)\.59,3)\.61,5)\.63,7)
\.65,9)\.66,10)\.67,11)\.56,2)\.57,3)\.60,6)\.61,7)\.62,8)\.63,9)\.64,10)
\.65,11)\.67,13)\.53,1)\.55,3)\.57,5)\.60,8)\.63,11)\.65,13)\.66,14)\.56,5)
\.59,8)\.62,11)\.56,6)\.58,8)\.60,10)\.63,13)\.65,15)\.51,3)\.52,4)\.53,5)
\.54,6)\.55,7)\.56,8)\.57,9)\.59,11)\.61,13)\.62,14)\.63,15)\.64,16)\.51,5)
\.52,6)\.55,9)\.53,8)\.54,9)\.58,13)\.59,14)\.62,17)\.48,4)\.49,5)\.52,8)
\.53,9)\.55,11)\.56,12)\.57,13)\.59,15)\.61,17)\.50,7)\.60,17)\.46,4)\.47,5)
\.48,6)\.49,7)\.50,8)\.51,9)\.53,11)\.54,12)\.55,13)\.56,14)\.57,15)\.60,18)
\.43,3)\.45,5)\.47,7)\.49,9)\.50,10)\.51,11)\.53,13)\.54,14)\.55,15)\.56,16)
\.57,17)\.59,19)\.50,11)\.56,18)\.38,2)\.40,4)\.41,5)\.42,6)\.43,7)\.44,8)
\.45,9)\.46,10)\.47,11)\.48,12)\.49,13)\.50,14)\.51,15)\.52,16)\.53,17)
\.55,19)\.56,20)\.57,21)\.58,22)\.43,8)\.44,9)\.54,19)\.38,4)\.44,10)\.46,12)
\.49,16)\.54,21)\.56,23)\.37,5)\.39,7)\.40,8)\.42,10)\.43,11)\.46,14)\.47,15)
\.49,17)\.52,20)\.54,22)\.55,23)\.56,24)\.36,6)\.39,9)\.41,11)\.42,12)\.43,13)
\.44,14)\.45,15)\.46,16)\.47,17)\.48,18)\.49,19)\.50,20)\.52,22)\.54,24)
\.36,7)\.39,10)\.33,5)\.37,9)\.39,11)\.41,13)\.45,17)\.47,19)\.48,20)\.50,22)
\.51,23)\.32,5)\.35,8)\.37,10)\.38,11)\.44,17)\.45,18)\.49,22)\.52,25)\.53,26)
\.45,19)\.39,14)\.49,24)\.29,5)\.31,7)\.33,9)\.34,10)\.35,11)\.36,12)\.37,13)
\.38,14)\.39,15)\.40,16)\.41,17)\.42,18)\.43,19)\.44,20)\.45,21)\.46,22)
\.47,23)\.48,24)\.49,25)\.50,26)\.51,27)\.52,28)\.33,11)\.50,28)\.51,29)
\.28,7)\.32,11)\.35,14)\.38,17)\.41,20)\.43,22)\.44,23)\.47,26)\.50,29)
\.27,7)\.28,8)\.29,9)\.32,12)\.33,13)\.34,14)\.35,15)\.36,16)\.37,17)\.39,19)
\.41,21)\.44,24)\.45,25)\.47,27)\.49,29)\.32,13)\.23,5)\.26,8)\.27,9)\.28,10)
\.29,11)\.32,14)\.33,15)\.34,16)\.35,17)\.36,18)\.38,20)\.40,22)\.42,24)
\.44,26)\.45,27)\.46,28)\.47,29)\.48,30)\.23,7)\.24,8)\.26,10)\.27,11)\.29,13)
\.30,14)\.31,15)\.32,16)\.33,17)\.35,19)\.36,20)\.39,23)\.41,25)\.43,27)
\.45,29)\.46,30)\.47,31)\.24,9)\.29,14)\.32,17)\.34,19)\.38,23)\.39,24)
\.42,27)\.43,28)\.44,29)\.47,32)\.25,11)\.31,17)\.34,20)\.38,24)\.43,29)
\.38,25)\.19,7)\.20,8)\.21,9)\.22,10)\.23,11)\.24,12)\.25,13)\.26,14)\.27,15)
\.28,16)\.29,17)\.30,18)\.31,19)\.32,20)\.33,21)\.34,22)\.35,23)\.36,24)
\.37,25)\.38,26)\.39,27)\.40,28)\.41,29)\.42,30)\.43,31)\.44,32)\.45,33)
\.46,34)\.26,15)\.18,8)\.20,10)\.25,15)\.27,17)\.29,19)\.33,23)\.34,24)
\.37,27)\.39,29)\.42,32)\.17,8)\.19,10)\.23,14)\.24,15)\.28,19)\.29,20)
\.32,23)\.35,26)\.36,27)\.37,28)\.44,35)\.18,10)\.19,11)\.22,14)\.23,15)
\.25,17)\.26,18)\.27,19)\.28,20)\.30,22)\.31,23)\.32,24)\.33,25)\.34,26)
\.35,27)\.36,28)\.37,29)\.39,31)\.40,32)\.41,33)\.42,34)\.43,35)\.27,20)
\.17,11)\.19,13)\.20,14)\.21,15)\.22,16)\.23,17)\.24,18)\.25,19)\.26,20)
\.27,21)\.28,22)\.29,23)\.30,24)\.31,25)\.32,26)\.35,29)\.36,30)\.38,32)
\.39,33)\.41,35)\.42,36)\.43,37)\.24,19)\.29,24)\.34,29)\.39,34)\.17,13)
\.19,15)\.20,16)\.21,17)\.24,20)\.27,23)\.28,24)\.29,25)\.30,26)\.31,27)
\.32,28)\.33,29)\.35,31)\.37,33)\.16,13)\.20,17)\.23,20)\.24,21)\.26,23)
\.29,26)\.32,29)\.35,32)\.38,35)\.40,37)\.20,18)\.28,26)\.33,31)\.34,32)
\.10,10)\.11,11)\.13,13)\.14,14)\.15,15)\.16,16)\.17,17)\.18,18)\.19,19)
\.20,20)\.21,21)\.22,22)\.23,23)\.24,24)\.25,25)\.26,26)\.27,27)\.28,28)
\.29,29)\.30,30)\.31,31)\.32,32)\.33,33)\.34,34)\.35,35)\.37,37)\.38,38)
\.39,39)\.12,13)\.21,23)\.24,26)\.26,28)\.32,34)\.33,35)\.13,16)\.15,18)
\.16,19)\.17,20)\.20,23)\.26,29)\.27,30)\.29,32)\.31,34)\.34,37)\.37,40)
\.15,19)\.16,20)\.18,22)\.19,23)\.20,24)\.21,25)\.23,27)\.24,28)\.25,29)
\.26,30)\.28,32)\.29,33)\.31,35)\.32,36)\.33,37)\.37,41)\.24,29)\.8,14)
\.10,16)\.12,18)\.13,19)\.14,20)\.15,21)\.16,22)\.18,24)\.19,25)\.21,27)
\.22,28)\.23,29)\.24,30)\.25,31)\.26,32)\.29,35)\.30,36)\.32,38)\.33,39)
\.35,41)\.37,43)\.19,26)\.27,34)\.9,17)\.13,21)\.14,22)\.15,23)\.17,25)
\.18,26)\.19,27)\.20,28)\.21,29)\.22,30)\.23,31)\.24,32)\.25,33)\.27,35)
\.28,36)\.29,37)\.33,41)\.34,42)\.35,43)\.18,27)\.19,28)\.20,29)\.23,32)
\.26,35)\.27,36)\.29,38)\.30,39)\.35,44)\.18,28)\.19,29)\.24,34)\.7,19)
\.8,20)\.9,21)\.10,22)\.11,23)\.12,24)\.13,25)\.14,26)\.15,27)\.16,28)\.17,29)
\.18,30)\.19,31)\.20,32)\.21,33)\.22,34)\.23,35)\.24,36)\.25,37)\.26,38)
\.27,39)\.28,40)\.29,41)\.30,42)\.31,43)\.32,44)\.34,46)\.12,26)\.20,34)
\.22,36)\.24,38)\.10,25)\.14,29)\.17,32)\.18,33)\.19,34)\.24,39)\.29,44)
\.31,46)\.32,47)\.7,23)\.11,27)\.13,29)\.15,31)\.17,33)\.18,34)\.19,35)
\.22,38)\.23,39)\.25,41)\.27,43)\.29,45)\.30,46)\.31,47)\.14,31)\.26,43)
\.6,24)\.8,26)\.9,27)\.10,28)\.11,29)\.12,30)\.14,32)\.16,34)\.17,35)\.18,36)
\.19,37)\.20,38)\.21,39)\.22,40)\.24,42)\.25,43)\.26,44)\.28,46)\.29,47)
\.30,48)\.5,25)\.9,29)\.12,32)\.13,33)\.14,34)\.15,35)\.16,36)\.17,37)\.18,38)
\.19,39)\.21,41)\.24,44)\.25,45)\.26,46)\.29,49)\.30,50)\.11,32)\.14,35)
\.16,37)\.17,38)\.20,41)\.29,50)\.11,33)\.15,37)\.17,39)\.18,40)\.21,43)
\.22,44)\.28,50)\.29,51)\.3,27)\.5,29)\.7,31)\.8,32)\.9,33)\.10,34)\.11,35)
\.12,36)\.13,37)\.14,38)\.15,39)\.16,40)\.17,41)\.18,42)\.19,43)\.20,44)
\.21,45)\.22,46)\.23,47)\.24,48)\.25,49)\.26,50)\.27,51)\.28,52)\.10,35)
\.19,44)\.19,45)\.7,34)\.8,35)\.10,37)\.11,38)\.15,42)\.16,43)\.17,44)\.23,50)
\.25,52)\.26,53)\.5,33)\.12,40)\.13,41)\.15,43)\.17,45)\.19,47)\.20,48)
\.21,49)\.22,50)\.23,51)\.3,33)\.5,35)\.6,36)\.7,37)\.8,38)\.9,39)\.11,41)
\.12,42)\.14,44)\.15,45)\.16,46)\.17,47)\.18,48)\.19,49)\.20,50)\.22,52)
\.24,54)\.9,41)\.11,43)\.14,46)\.15,47)\.17,49)\.19,51)\.22,54)\.24,56)
\.14,47)\.15,48)\.16,49)\.18,51)\.20,53)\.21,54)\.23,56)\.15,49)\.22,56)
\.20,55)\.2,38)\.4,40)\.5,41)\.6,42)\.7,43)\.8,44)\.9,45)\.10,46)\.11,47)
\.12,48)\.13,49)\.14,50)\.15,51)\.16,52)\.17,53)\.19,55)\.20,56)\.21,57)
\.22,58)\.15,53)\.11,50)\.20,59)\.7,47)\.9,49)\.10,50)\.11,51)\.13,53)\.14,54)
\.16,56)\.17,57)\.18,58)\.19,59)\.4,46)\.5,47)\.6,48)\.8,50)\.11,53)\.12,54)
\.13,55)\.14,56)\.15,57)\.16,58)\.18,60)\.19,61)\.7,50)\.5,49)\.8,52)\.9,53)
\.12,56)\.13,57)\.4,49)\.9,54)\.10,55)\.11,56)\.13,58)\.14,59)\.17,62)\.1,49)
\.3,51)\.5,53)\.6,54)\.7,55)\.8,56)\.9,57)\.11,59)\.13,61)\.14,62)\.15,63)
\.16,64)\.6,56)\.10,60)\.15,65)\.5,56)\.8,59)\.3,55)\.5,57)\.8,60)\.11,63)
\.12,64)\.13,65)\.14,66)\.8,61)\.2,56)\.6,60)\.10,64)\.11,65)\.12,66)\.13,67)
\.9,64)\.3,59)\.5,61)\.7,63)\.10,66)\.11,67)\.12,68)\.11,68)\.7,65)\.1,61)
\.2,62)\.3,63)\.4,64)\.5,65)\.6,66)\.7,67)\.8,68)\.9,69)\.10,70)\.3,66)
\.7,70)\.8,71)\.5,69)\.7,71)\.3,69)\.5,71)\.6,72)\.5,73)\.5,74)\.1,73)\.2,74)
\.3,75)\.4,76)\.1,77) % #spec=846
}

% ========================     Fig. 3     ========================== %
\def\figsca{\funit=0.5truemm \wi=330 \he=180 \vmul=1 }
\def\figlab{\hlab{-200}\hlab{200} \hlab{-100}\hlab{100} \hlab{-300}\hlab{300}
	    \vlab{100}  \vlab{150} \vlab{50}}
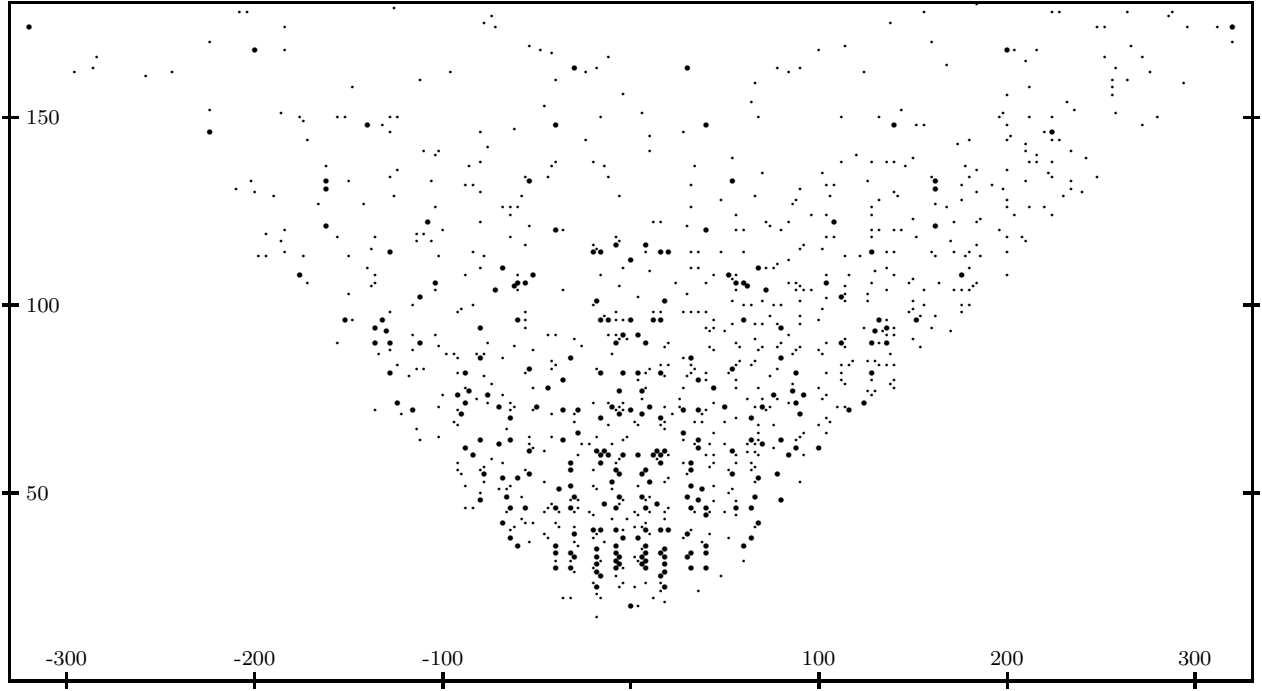
\def\figcap{$\ng+\na$ vs. Euler number $\chi$ for the 258 spectra
     that do have a mirror spectrum\\ ~~~~~~~~but cannot be obtained from an
     invertible skeleton (the ones without mirror and \\
     with $\ng+\na\le180$ are indicated by small dots)\hskip 45mm~  }

\def\nilehe{     % non-invertible, no mirror, g+a <= (\he=180)
\.155,7)\.153,10)\.154,12)\.145,16)\.142,20)\.132,20)\.141,29)\.118,13)
\.141,37)\.140,38)\.117,16)\.115,15)\.106,7)\.105,8)\.108,11)\.112,17)\.105,12)
\.122,29)\.103,11)\.106,14)\.130,38)\.133,41)\.119,31)\.100,13)\.118,31)
\.96,10)\.115,29)\.105,22)\.109,28)\.84,6)\.98,20)\.114,36)\.113,37)\.89,14)
\.94,19)\.104,29)\.85,11)\.116,42)\.99,28)\.90,20)\.87,18)\.92,23)\.70,2)
\.87,19)\.88,20)\.93,25)\.77,10)\.107,41)\.76,12)\.81,17)\.99,35)\.105,41)
\.107,43)\.96,33)\.121,58)\.73,11)\.99,37)\.106,44)\.66,5)\.70,11)\.73,15)
\.75,17)\.79,21)\.62,5)\.70,13)\.87,30)\.88,31)\.60,4)\.108,52)\.69,14)
\.87,32)\.98,43)\.85,32)\.93,40)\.72,20)\.78,26)\.89,37)\.96,44)\.58,7)
\.62,11)\.70,19)\.82,31)\.96,45)\.63,13)\.85,35)\.68,19)\.105,57)\.58,11)
\.67,20)\.51,5)\.52,6)\.57,11)\.59,13)\.64,18)\.66,20)\.72,26)\.50,5)\.72,27)
\.45,1)\.48,4)\.61,17)\.76,32)\.88,44)\.90,47)\.44,2)\.71,29)\.78,36)\.87,45)
\.58,17)\.64,23)\.45,5)\.48,8)\.75,35)\.81,41)\.85,45)\.89,49)\.93,53)\.46,7)
\.48,9)\.60,21)\.63,24)\.76,37)\.107,68)\.56,18)\.64,26)\.90,52)\.48,11)
\.58,21)\.107,70)\.105,69)\.43,8)\.46,12)\.55,21)\.60,26)\.80,46)\.39,6)
\.42,9)\.50,17)\.36,4)\.42,10)\.52,20)\.53,21)\.75,43)\.78,46)\.79,47)\.36,5)
\.52,21)\.60,29)\.89,58)\.67,37)\.69,39)\.78,48)\.36,7)\.39,10)\.79,50)
\.35,7)\.42,14)\.55,27)\.61,33)\.62,34)\.63,35)\.72,44)\.81,53)\.32,5)\.45,18)
\.46,19)\.74,47)\.98,71)\.34,8)\.44,18)\.49,23)\.52,26)\.57,31)\.96,72)
\.31,8)\.34,11)\.38,15)\.42,19)\.88,65)\.30,8)\.34,12)\.57,35)\.78,56)\.34,13)
\.43,22)\.46,25)\.52,31)\.53,32)\.79,58)\.94,73)\.26,6)\.31,11)\.52,32)
\.56,36)\.79,59)\.90,70)\.30,11)\.32,13)\.20,2)\.61,43)\.69,51)\.19,3)\.24,8)
\.32,16)\.22,7)\.26,11)\.28,13)\.30,15)\.42,27)\.43,28)\.53,38)\.25,11)
\.30,16)\.31,17)\.37,23)\.38,25)\.51,38)\.78,66)\.87,75)\.28,17)\.37,26)
\.18,8)\.20,10)\.24,14)\.27,17)\.31,21)\.33,23)\.42,32)\.55,45)\.63,53)
\.74,64)\.13,4)\.16,7)\.24,15)\.30,21)\.37,28)\.46,37)\.62,53)\.86,77)\.15,7)
\.20,12)\.58,50)\.79,71)\.22,15)\.52,45)\.72,65)\.37,31)\.40,34)\.49,43)
\.70,64)\.86,80)\.24,19)\.28,23)\.34,29)\.46,41)\.16,12)\.31,27)\.38,34)
\.48,44)\.50,46)\.51,47)\.61,57)\.14,11)\.25,22)\.31,28)\.38,35)\.47,44)
\.54,51)\.60,57)\.66,63)\.14,12)\.16,14)\.21,19)\.25,23)\.33,31)\.37,35)
\.38,36)\.49,47)\.56,54)\.79,77)\.22,21)\.33,32)\.46,45)\.12,13)\.16,17)
\.20,21)\.22,23)\.34,35)\.9,11)\.15,17)\.21,23)\.25,27)\.33,35)\.39,41)
\.50,52)\.16,19)\.19,22)\.36,39)\.74,77)\.11,15)\.19,23)\.22,26)\.37,41)
\.39,43)\.55,59)\.17,22)\.20,25)\.22,27)\.39,44)\.52,57)\.68,73)\.70,75)
\.8,14)\.46,52)\.58,64)\.26,33)\.8,16)\.9,17)\.30,38)\.38,46)\.45,53)\.57,65)
\.6,15)\.17,26)\.18,27)\.22,31)\.29,38)\.30,39)\.32,41)\.36,45)\.40,49)
\.18,28)\.22,32)\.25,35)\.37,47)\.40,50)\.30,41)\.47,59)\.30,43)\.31,44)
\.12,26)\.22,36)\.25,39)\.32,46)\.36,50)\.42,56)\.66,80)\.20,35)\.33,48)
\.44,59)\.53,68)\.12,28)\.33,49)\.34,50)\.41,57)\.47,63)\.49,65)\.55,71)
\.58,74)\.14,31)\.23,40)\.26,43)\.33,50)\.60,77)\.3,21)\.40,58)\.46,64)
\.48,66)\.9,28)\.25,44)\.26,45)\.20,40)\.26,46)\.30,50)\.31,51)\.37,57)
\.39,59)\.43,63)\.9,30)\.12,33)\.18,39)\.21,42)\.34,55)\.36,57)\.13,35)
\.15,37)\.18,40)\.19,41)\.22,44)\.26,48)\.33,55)\.36,58)\.37,59)\.43,65)
\.2,26)\.45,69)\.52,76)\.10,35)\.19,44)\.22,47)\.34,59)\.6,32)\.12,38)\.15,41)
\.28,54)\.12,39)\.34,61)\.37,64)\.40,67)\.56,83)\.8,36)\.10,38)\.15,43)
\.16,44)\.21,49)\.25,53)\.26,54)\.28,56)\.31,59)\.38,66)\.40,68)\.46,74)
\.48,76)\.28,57)\.30,59)\.1,31)\.35,65)\.42,72)\.45,75)\.15,46)\.16,47)
\.22,53)\.12,44)\.13,45)\.26,58)\.34,66)\.61,93)\.4,37)\.10,43)\.12,45)
\.14,47)\.15,48)\.20,53)\.26,59)\.29,62)\.31,64)\.48,81)\.63,96)\.13,47)
\.15,49)\.16,50)\.17,51)\.22,56)\.27,61)\.58,92)\.20,55)\.27,62)\.30,65)
\.33,68)\.50,85)\.18,54)\.37,73)\.43,79)\.18,55)\.22,59)\.30,67)\.42,79)
\.12,50)\.15,53)\.31,69)\.36,74)\.18,57)\.22,61)\.25,64)\.28,67)\.36,75)
\.62,101)\.25,65)\.30,70)\.40,80)\.16,58)\.25,67)\.32,74)\.35,77)\.40,82)
\.60,102)\.18,61)\.22,65)\.42,85)\.10,54)\.22,66)\.30,74)\.31,75)\.42,86)
\.4,49)\.10,55)\.22,67)\.25,70)\.26,71)\.28,73)\.30,75)\.35,80)\.43,88)
\.59,104)\.7,53)\.10,56)\.29,75)\.35,81)\.28,76)\.38,86)\.9,59)\.14,64)
\.19,69)\.23,73)\.59,109)\.12,63)\.18,69)\.24,75)\.32,83)\.42,93)\.7,59)
\.9,61)\.12,64)\.15,67)\.19,71)\.25,77)\.28,80)\.33,85)\.36,88)\.40,92)
\.8,61)\.38,91)\.32,86)\.39,93)\.9,64)\.12,68)\.13,69)\.14,70)\.41,97)\.47,103)
\.8,65)\.11,68)\.16,73)\.17,74)\.22,79)\.23,80)\.24,81)\.28,85)\.56,113)
\.13,71)\.25,83)\.13,72)\.40,100)\.26,87)\.16,78)\.50,112)\.7,70)\.11,74)
\.19,82)\.20,83)\.35,98)\.6,70)\.8,72)\.21,85)\.27,91)\.28,92)\.30,94)\.31,95)
\.37,101)\.6,71)\.10,75)\.22,87)\.9,75)\.12,78)\.27,93)\.31,97)\.42,108)
\.13,80)\.8,76)\.12,80)\.14,82)\.18,86)\.21,89)\.23,91)\.35,103)\.5,74)
\.9,78)\.53,122)\.4,74)\.5,75)\.7,77)\.11,81)\.12,82)\.19,89)\.22,92)\.25,95)
\.30,100)\.34,104)\.29,100)\.24,96)\.40,112)\.32,105)\.12,86)\.16,90)\.8,83)
\.12,87)\.13,88)\.16,91)\.24,99)\.32,107)\.21,97)\.36,112)\.6,83)\.10,87)
\.11,89)\.13,91)\.35,113)\.50,128)\.7,87)\.17,97)\.27,107)\.45,125)\.8,89)
\.16,97)\.23,104)\.12,94)\.21,105)\.40,124)\.4,89)\.16,101)\.9,95)\.14,100)
\.18,104)\.28,115)\.5,93)\.6,94)\.8,96)\.20,108)\.22,110)\.12,101)\.24,113)
\.5,95)\.7,97)\.10,100)\.12,102)\.22,112)\.27,117)\.7,98)\.7,99)\.12,104)
\.15,107)\.19,111)\.22,114)\.44,136)\.10,103)\.15,108)\.18,114)\.11,109)
\.26,124)\.16,115)\.26,125)\.4,104)\.7,107)\.9,109)\.13,113)\.14,114)\.18,118)
\.22,122)\.24,124)\.28,128)\.22,123)\.11,113)\.33,135)\.7,111)\.6,111)\.14,119)
\.18,123)\.19,124)\.30,135)\.10,116)\.16,122)\.22,128)\.26,132)\.15,123)
\.16,124)\.30,138)\.8,118)\.18,128)\.12,123)\.6,118)\.11,123)\.13,125)\.16,128)
\.19,131)\.33,145)\.15,129)\.32,146)\.7,122)\.19,135)\.7,124)\.8,125)\.12,129)
\.17,135)\.5,125)\.9,130)\.5,129)\.25,149)\.20,146)\.24,150)\.14,142)\.15,143)
\.16,144)\.11,140)\.17,146)\.14,146)\.23,155)\.18,151)\.6,142)\.15,151)
\.12,150)\.5,145)\.17,160)\.17,161)\.6,153)\.13,161)\.9,165)\.5,165) %#spec=727
}
\Vplo{   \def\npt{\funit=1pt\circle*2}
\.167,7)\.129,17)\.134,34)\.98,10)\.101,20)\.106,25)\.107,26)\.86,10)\.109,39)
\.79,11)\.81,13)\.81,15)\.79,14)\.73,9)\.77,13)\.89,25)\.68,6)\.65,7)\.73,17)
\.79,23)\.88,34)\.79,27)\.61,15)\.58,13)\.53,9)\.59,15)\.63,19)\.60,17)
\.51,9)\.44,4)\.52,12)\.63,23)\.67,27)\.47,8)\.57,19)\.70,34)\.49,14)\.54,19)
\.38,4)\.44,10)\.72,38)\.41,8)\.35,3)\.39,7)\.48,16)\.51,19)\.68,37)\.33,3)
\.42,12)\.63,33)\.68,38)\.37,9)\.67,39)\.41,14)\.44,17)\.55,28)\.80,53)
\.67,41)\.49,24)\.50,28)\.25,5)\.27,7)\.28,8)\.33,13)\.70,50)\.84,64)\.35,16)
\.41,23)\.45,27)\.49,31)\.23,7)\.25,9)\.31,15)\.34,18)\.36,20)\.37,21)\.51,35)
\.24,9)\.27,12)\.32,17)\.89,74)\.40,26)\.43,29)\.25,15)\.62,52)\.17,8)\.19,10)
\.20,11)\.21,12)\.22,13)\.35,26)\.55,46)\.18,10)\.24,16)\.33,25)\.34,26)
\.39,31)\.45,37)\.52,44)\.61,53)\.27,20)\.34,27)\.33,27)\.51,45)\.29,24)
\.39,34)\.17,13)\.18,14)\.19,15)\.20,16)\.22,18)\.25,21)\.30,26)\.47,43)
\.60,56)\.17,14)\.18,15)\.26,23)\.29,26)\.37,34)\.40,37)\.20,18)\.31,29)
\.42,40)\.47,45)\.10,10)\.36,36)\.48,48)\.56,56)\.18,20)\.29,31)\.40,42)
\.45,47)\.14,17)\.15,18)\.23,26)\.26,29)\.34,37)\.37,40)\.13,17)\.14,18)
\.15,19)\.16,20)\.18,22)\.21,25)\.26,30)\.43,47)\.56,60)\.24,29)\.34,39)
\.27,33)\.45,51)\.20,27)\.27,34)\.10,18)\.13,21)\.16,24)\.25,33)\.26,34)
\.31,39)\.37,45)\.44,52)\.53,61)\.8,17)\.10,19)\.11,20)\.12,21)\.13,22)
\.26,35)\.46,55)\.15,25)\.52,62)\.26,40)\.29,43)\.9,24)\.12,27)\.17,32)
\.74,89)\.7,23)\.9,25)\.15,31)\.18,34)\.20,36)\.21,37)\.35,51)\.15,33)\.22,40)
\.23,41)\.27,45)\.31,49)\.16,35)\.5,25)\.7,27)\.8,28)\.12,32)\.13,33)\.50,70)
\.64,84)\.28,50)\.24,49)\.41,67)\.14,41)\.17,44)\.28,55)\.53,80)\.9,37)
\.39,67)\.3,33)\.33,63)\.38,68)\.37,68)\.3,35)\.7,39)\.16,48)\.19,51)\.8,41)
\.4,38)\.10,44)\.38,72)\.14,49)\.19,54)\.34,70)\.19,57)\.8,47)\.4,44)\.12,52)
\.23,63)\.27,67)\.9,51)\.17,60)\.9,53)\.15,59)\.19,63)\.13,58)\.15,61)\.6,56)
\.27,79)\.34,88)\.17,73)\.23,79)\.7,65)\.6,68)\.9,73)\.13,77)\.25,89)\.14,79)
\.15,81)\.11,79)\.13,81)\.39,109)\.10,86)\.20,101)\.25,106)\.26,107)\.10,98)
\.34,134)\.17,129)\.7,167) % #spec=258
\def\npt{\funit=1pt\circle*1} \nilehe
}
\newpage

% ========================     Fig. 4     ========================== %
\def\figsca{\funit=0.15truemm \wi=900 \he=450 \vmul=2 }
\def\figlab{\hlab{-200} \hlab{200} \hlab{-400} \hlab{400}
            \hlab{-600} \hlab{600} \hlab{-800} \hlab{800}
	    \vlab{100}  \vlab{200} \vlab{300}  \vlab{400} }
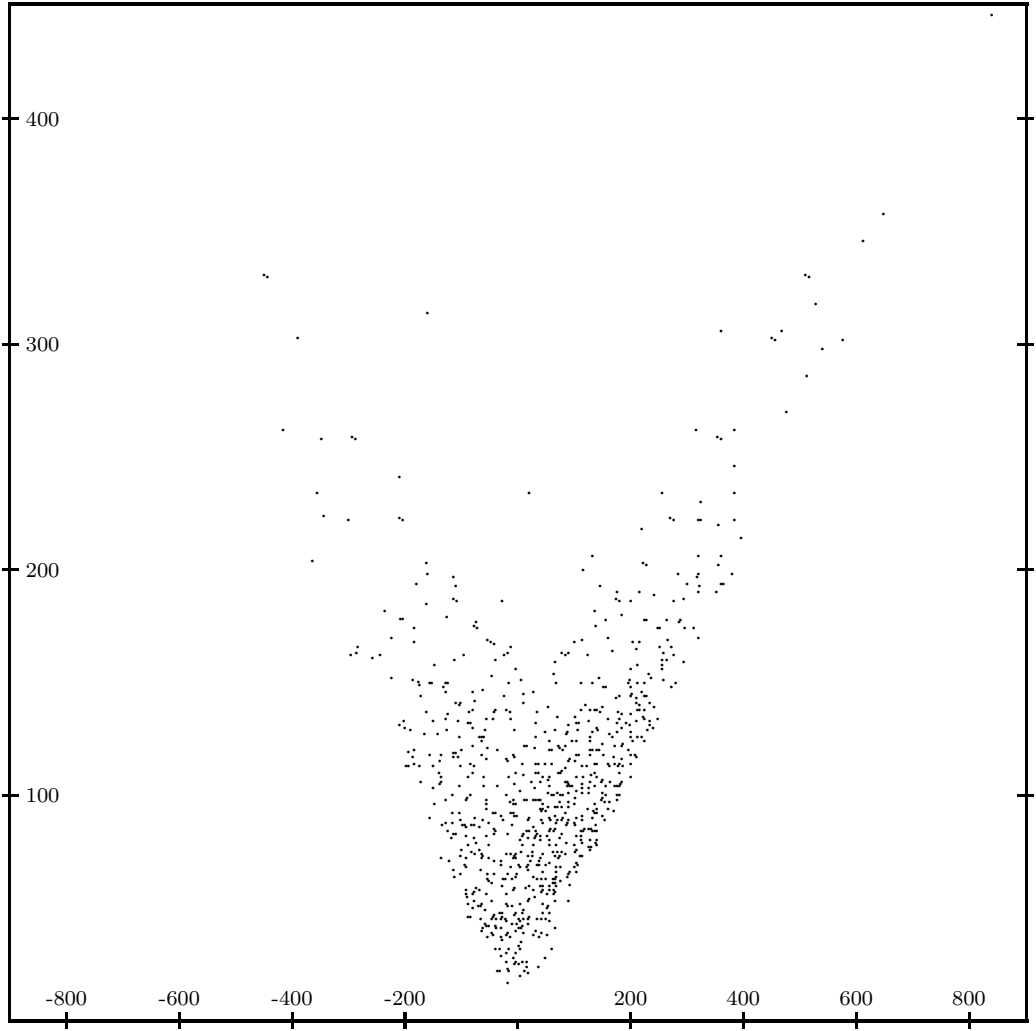
\def\figcap{$\ng+\na$ vs. Euler number $\chi$ for the 810 spectra without
            mirror}
\Vplo{   \nilehe  \def\nilehe{}
\.278,53)\.276,54)\.235,27)\.249,54)\.193,11)\.206,28)\.216,42)\.198,26)
\.186,36)\.203,56)\.201,57)\.150,32)\.164,59)\.173,68)\.162,60)\.142,52)
\.133,52)\.142,61)\.139,59)\.197,117)\.122,65)\.127,70)\.124,69)\.120,66)
\.100,86)\.112,122)\.71,129)\.70,136)\.57,125)\.60,133)\.50,137)\.51,139)
\.48,138)\.43,143)\.41,149)\.54,164)\.46,157)\.44,158)\.34,155)\.53,181)
\.44,179)\.24,162)\.42,180)\.28,170)\.20,167)\.22,172)\.52,210)\.19,178)
\.15,175)\.19,179)\.23,183)\.31,191)\.16,177)\.30,192)\.34,196)\.7,183)
\.41,218)\.12,190)\.21,199)\.7,187)\.13,193)\.39,219)\.63,243)\.6,188)\.4,194)
\.15,207)\.21,213)\.27,219)\.35,227)\.8,206)\.39,264)\.37,265)\.36,270)
\.16,254)\.38,293)\.15,271)\.36,294)\.27,291)\.14,284)\.7,295)\.20,326)
\.17,341)\.13,433) % #spec=83
}
\newpage

% ========================     Fig. 5     ========================== %
\def\figsca{\funit=0.15truemm \wi=1000 \he=520 \vmul=2 }
\def\figlab{\hlab{-200} \hlab{200} \hlab{-400} \hlab{400}
	    \hlab{-600} \hlab{600} \hlab{-800} \hlab{800}
	    \vlab{100}  \vlab{200} \vlab{300}  \vlab{400} \vmark{500} }
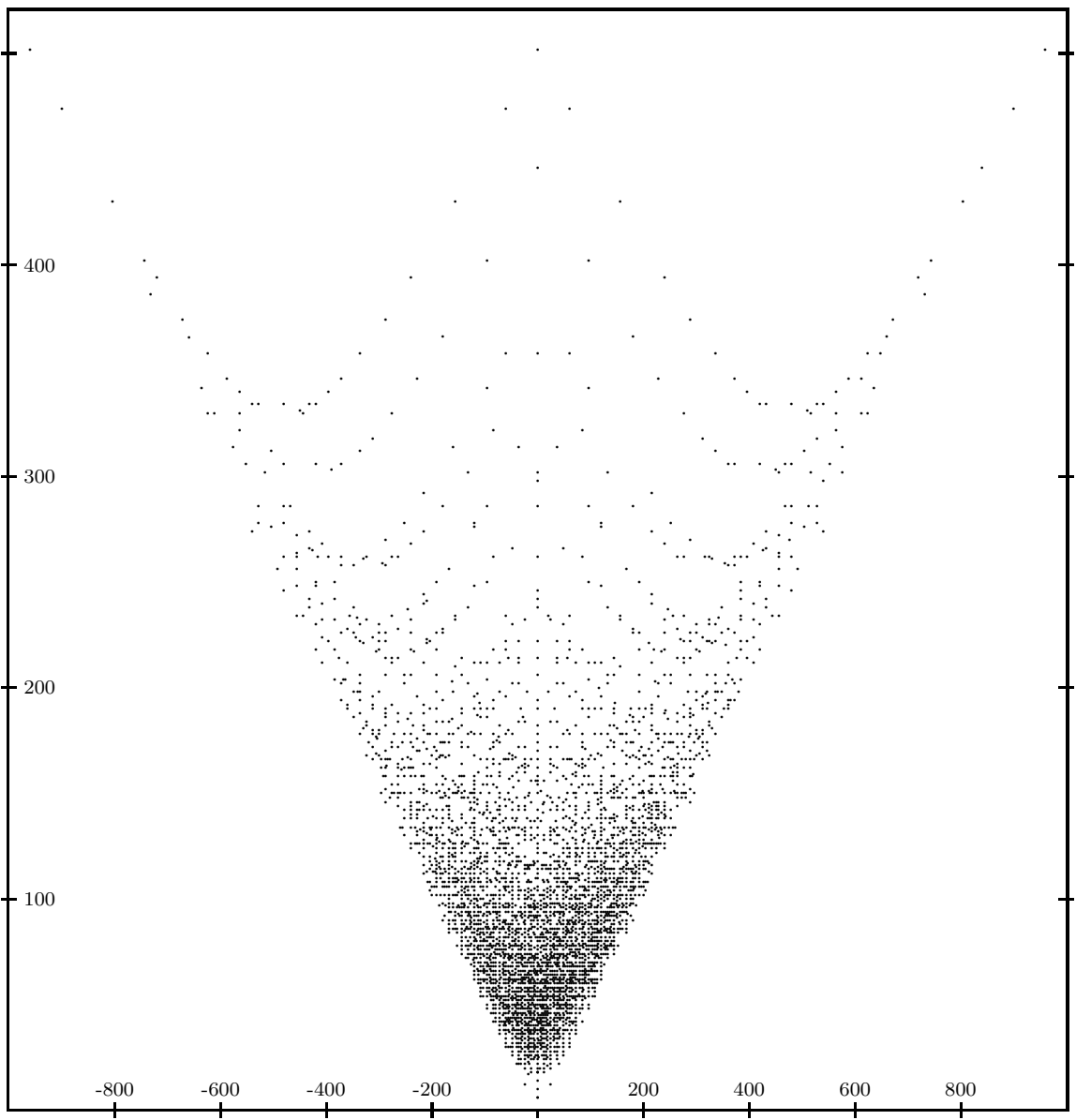
\def\figcap{$\ng+\na$ vs. Euler number $\chi$ for all 3837 different spectra}
\Vplo{
\.491,11)\.462,12)\.416,14)\.387,15)\.376,10)\.377,17)\.355,19)\.348,18)
\.330,12)\.321,9)\.335,23)\.318,12)\.320,26)\.301,13)\.302,20)\.306,24)
\.311,29)\.291,15)\.272,2)\.302,32)\.271,7)\.275,11)\.299,35)\.280,22)\.264,12)
\.282,30)\.251,5)\.243,3)\.251,11)\.259,19)\.263,23)\.273,33)\.287,47)\.260,26)
\.231,3)\.238,10)\.242,14)\.245,17)\.246,18)\.250,22)\.278,53)\.228,6)\.276,54)
\.227,11)\.229,13)\.241,25)\.245,29)\.275,59)\.239,26)\.214,4)\.220,10)
\.229,19)\.230,20)\.258,48)\.272,62)\.235,27)\.208,4)\.214,10)\.218,14)
\.222,18)\.236,32)\.210,12)\.230,32)\.269,71)\.249,54)\.198,6)\.205,13)
\.212,20)\.217,25)\.221,29)\.201,13)\.190,4)\.194,8)\.206,20)\.222,36)\.224,38)
\.246,60)\.266,80)\.194,10)\.193,11)\.185,5)\.186,6)\.196,16)\.204,24)\.206,28)
\.180,6)\.186,12)\.200,26)\.206,32)\.216,42)\.198,26)\.202,31)\.184,14)
\.173,5)\.178,10)\.181,13)\.183,15)\.185,17)\.187,19)\.195,27)\.199,31)
\.240,72)\.263,95)\.173,8)\.193,28)\.213,48)\.165,3)\.172,10)\.173,11)\.197,35)
\.212,50)\.167,7)\.161,5)\.164,8)\.167,11)\.168,12)\.173,17)\.174,18)\.176,20)
\.190,34)\.237,81)\.161,8)\.165,12)\.185,32)\.164,12)\.159,9)\.164,14)\.165,15)
\.174,24)\.184,34)\.186,36)\.188,38)\.190,40)\.149,1)\.155,7)\.157,9)\.203,56)
\.152,6)\.145,1)\.148,4)\.151,7)\.153,9)\.155,11)\.157,13)\.163,19)\.165,21)
\.166,22)\.167,23)\.175,31)\.179,35)\.183,39)\.185,41)\.201,57)\.207,63)
\.259,115)\.153,10)\.154,12)\.144,4)\.153,13)\.144,6)\.146,8)\.155,17)\.156,18)
\.165,27)\.175,37)\.176,38)\.186,48)\.200,62)\.234,96)\.143,7)\.157,21)
\.138,6)\.140,8)\.141,9)\.143,11)\.145,13)\.147,15)\.148,16)\.152,20)\.158,26)
\.160,28)\.173,41)\.180,48)\.197,65)\.132,2)\.140,10)\.145,16)\.131,3)\.147,19)
\.151,23)\.128,2)\.132,6)\.134,8)\.138,12)\.142,16)\.144,18)\.152,26)\.164,38)
\.166,40)\.202,76)\.145,22)\.154,31)\.180,57)\.142,20)\.122,2)\.123,3)\.127,7)
\.129,9)\.130,10)\.131,11)\.132,12)\.133,13)\.135,15)\.139,19)\.141,21)
\.143,23)\.147,27)\.149,29)\.154,34)\.163,43)\.169,49)\.173,53)\.176,56)
\.194,74)\.257,137)\.140,22)\.145,27)\.150,32)\.167,50)\.121,5)\.125,9)
\.137,21)\.120,6)\.122,8)\.123,9)\.126,12)\.127,13)\.128,14)\.130,16)\.142,28)
\.145,31)\.155,41)\.230,116)\.117,5)\.119,7)\.127,15)\.129,17)\.132,20)
\.141,29)\.122,11)\.134,24)\.110,2)\.112,4)\.114,6)\.116,8)\.117,9)\.118,10)
\.120,12)\.121,13)\.122,14)\.125,17)\.126,18)\.129,21)\.130,22)\.131,23)
\.133,25)\.136,28)\.141,33)\.144,36)\.145,37)\.148,40)\.150,42)\.160,52)
\.174,66)\.176,68)\.191,83)\.200,92)\.109,4)\.118,13)\.125,20)\.128,23)
\.138,33)\.163,58)\.164,59)\.173,68)\.106,2)\.115,11)\.121,17)\.141,37)
\.103,1)\.105,3)\.108,6)\.110,8)\.111,9)\.114,12)\.116,14)\.118,16)\.122,20)
\.133,31)\.140,38)\.146,44)\.162,60)\.117,16)\.101,1)\.103,3)\.108,8)\.115,15)
\.117,17)\.125,25)\.134,34)\.106,7)\.112,13)\.119,20)\.102,4)\.135,37)\.105,8)
\.108,11)\.99,3)\.101,5)\.102,6)\.103,7)\.104,8)\.105,9)\.106,10)\.107,11)
\.109,13)\.110,14)\.111,15)\.112,16)\.113,17)\.115,19)\.117,21)\.119,23)
\.123,27)\.125,29)\.129,33)\.131,35)\.133,37)\.137,41)\.140,44)\.143,47)
\.147,51)\.151,55)\.157,61)\.161,65)\.173,77)\.112,17)\.95,2)\.104,11)\.105,12)
\.110,17)\.122,29)\.130,37)\.97,5)\.101,9)\.102,10)\.103,11)\.106,14)\.113,21)
\.120,28)\.130,38)\.133,41)\.90,0)\.94,4)\.98,8)\.99,9)\.101,11)\.104,14)
\.105,15)\.107,17)\.108,18)\.110,20)\.113,23)\.114,24)\.116,26)\.119,29)
\.122,32)\.124,34)\.129,39)\.132,42)\.137,47)\.142,52)\.158,68)\.159,69)
\.188,98)\.228,138)\.90,2)\.95,7)\.97,9)\.98,10)\.101,13)\.107,19)\.119,31)
\.131,43)\.92,5)\.100,13)\.118,31)\.122,35)\.127,40)\.96,10)\.115,29)\.84,0)
\.85,1)\.86,2)\.87,3)\.89,5)\.90,6)\.91,7)\.92,8)\.93,9)\.95,11)\.96,12)
\.97,13)\.98,14)\.99,15)\.100,16)\.101,17)\.102,18)\.103,19)\.104,20)\.106,22)
\.109,25)\.111,27)\.112,28)\.113,29)\.114,30)\.116,32)\.118,34)\.126,42)
\.129,45)\.131,47)\.134,50)\.170,86)\.105,22)\.88,6)\.98,17)\.101,20)\.106,25)
\.107,26)\.109,28)\.133,52)\.142,61)\.83,3)\.85,5)\.88,8)\.89,9)\.91,11)
\.94,14)\.95,15)\.100,20)\.102,22)\.103,23)\.115,35)\.118,38)\.129,49)\.139,59)
\.197,117)\.82,4)\.83,5)\.84,6)\.85,7)\.86,8)\.90,12)\.92,14)\.93,15)\.94,16)
\.95,17)\.96,18)\.98,20)\.99,21)\.100,22)\.101,23)\.102,24)\.105,27)\.114,36)
\.116,38)\.144,66)\.155,77)\.156,78)\.254,176)\.77,1)\.80,4)\.81,5)\.86,10)
\.89,13)\.92,16)\.93,17)\.103,27)\.105,29)\.113,37)\.84,9)\.86,11)\.89,14)
\.92,17)\.94,19)\.104,29)\.113,38)\.114,39)\.128,53)\.83,9)\.85,11)\.116,42)
\.73,1)\.74,2)\.75,3)\.76,4)\.77,5)\.78,6)\.79,7)\.80,8)\.81,9)\.82,10)
\.83,11)\.84,12)\.85,13)\.86,14)\.87,15)\.88,16)\.89,17)\.90,18)\.91,19)
\.92,20)\.93,21)\.95,23)\.97,25)\.98,26)\.99,27)\.101,29)\.103,31)\.105,33)
\.106,34)\.110,38)\.111,39)\.113,41)\.115,43)\.118,46)\.122,50)\.123,51)
\.127,55)\.130,58)\.131,59)\.139,67)\.143,71)\.99,28)\.79,9)\.83,13)\.84,14)
\.90,20)\.109,39)\.114,44)\.77,8)\.83,14)\.86,17)\.87,18)\.92,23)\.97,28)
\.110,41)\.70,2)\.73,5)\.75,7)\.77,9)\.79,11)\.81,13)\.85,17)\.87,19)\.88,20)
\.90,22)\.93,25)\.97,29)\.102,34)\.77,10)\.69,3)\.70,4)\.71,5)\.72,6)\.73,7)
\.76,10)\.79,13)\.80,14)\.81,15)\.83,17)\.84,18)\.85,19)\.86,20)\.90,24)
\.96,30)\.107,41)\.119,53)\.122,56)\.134,68)\.152,86)\.184,118)\.79,14)
\.69,5)\.71,7)\.73,9)\.75,11)\.76,12)\.77,13)\.78,14)\.79,15)\.81,17)\.83,19)
\.86,22)\.87,23)\.88,24)\.89,25)\.93,29)\.99,35)\.103,39)\.105,41)\.107,43)
\.66,3)\.71,8)\.75,12)\.76,13)\.77,14)\.81,18)\.86,23)\.89,26)\.90,27)\.96,33)
\.103,40)\.121,58)\.68,6)\.73,11)\.84,22)\.99,37)\.106,44)\.66,5)\.61,1)
\.62,2)\.63,3)\.64,4)\.65,5)\.66,6)\.67,7)\.68,8)\.69,9)\.70,10)\.71,11)
\.72,12)\.73,13)\.74,14)\.75,15)\.76,16)\.77,17)\.78,18)\.79,19)\.80,20)
\.81,21)\.82,22)\.83,23)\.84,24)\.85,25)\.86,26)\.87,27)\.88,28)\.89,29)
\.91,31)\.92,32)\.94,34)\.95,35)\.96,36)\.97,37)\.98,38)\.99,39)\.101,41)
\.103,43)\.104,44)\.105,45)\.106,46)\.107,47)\.108,48)\.109,49)\.114,54)
\.119,59)\.125,65)\.128,68)\.136,76)\.154,94)\.168,108)\.169,109)\.70,11)
\.65,7)\.69,11)\.73,15)\.75,17)\.79,21)\.91,33)\.62,5)\.67,10)\.70,13)\.71,14)
\.77,20)\.78,21)\.86,29)\.87,30)\.88,31)\.92,35)\.101,44)\.102,45)\.122,65)
\.127,70)\.58,2)\.59,3)\.60,4)\.61,5)\.63,7)\.65,9)\.66,10)\.67,11)\.71,15)
\.72,16)\.73,17)\.74,18)\.75,19)\.76,20)\.79,23)\.82,26)\.83,27)\.87,31)
\.90,34)\.91,35)\.108,52)\.69,14)\.87,32)\.98,43)\.124,69)\.54,0)\.55,1)
\.56,2)\.57,3)\.58,4)\.60,6)\.61,7)\.62,8)\.63,9)\.64,10)\.65,11)\.66,12)
\.67,13)\.68,14)\.69,15)\.70,16)\.71,17)\.72,18)\.74,20)\.75,21)\.77,23)
\.80,26)\.81,27)\.83,29)\.84,30)\.87,33)\.88,34)\.92,38)\.102,48)\.110,56)
\.116,62)\.120,66)\.122,68)\.133,79)\.85,32)\.93,40)\.53,1)\.55,3)\.57,5)
\.59,7)\.60,8)\.63,11)\.65,13)\.66,14)\.68,16)\.69,17)\.72,20)\.76,24)\.78,26)
\.79,27)\.89,37)\.96,44)\.103,51)\.56,5)\.58,7)\.59,8)\.62,11)\.66,15)\.67,16)
\.70,19)\.72,21)\.74,23)\.82,31)\.96,45)\.56,6)\.58,8)\.60,10)\.63,13)\.65,15)
\.74,24)\.84,34)\.85,35)\.68,19)\.48,0)\.49,1)\.51,3)\.52,4)\.53,5)\.54,6)
\.55,7)\.56,8)\.57,9)\.58,10)\.59,11)\.60,12)\.61,13)\.62,14)\.63,15)\.64,16)
\.65,17)\.66,18)\.67,19)\.68,20)\.69,21)\.70,22)\.71,23)\.72,24)\.73,25)
\.75,27)\.77,29)\.78,30)\.79,31)\.80,32)\.81,33)\.83,35)\.84,36)\.85,37)
\.87,39)\.91,43)\.93,45)\.95,47)\.96,48)\.97,49)\.101,53)\.103,55)\.105,57)
\.107,59)\.117,69)\.119,71)\.126,78)\.130,82)\.149,101)\.167,119)\.195,147)
\.225,177)\.58,11)\.67,20)\.51,5)\.52,6)\.55,9)\.57,11)\.59,13)\.61,15)
\.62,16)\.64,18)\.66,20)\.70,24)\.72,26)\.98,52)\.50,5)\.53,8)\.54,9)\.56,11)
\.58,13)\.59,14)\.62,17)\.64,19)\.69,24)\.72,27)\.77,32)\.89,44)\.93,48)
\.108,63)\.45,1)\.48,4)\.49,5)\.52,8)\.53,9)\.55,11)\.56,12)\.57,13)\.59,15)
\.61,17)\.63,19)\.64,20)\.73,29)\.76,32)\.88,44)\.101,57)\.50,7)\.60,17)
\.90,47)\.42,0)\.44,2)\.45,3)\.46,4)\.47,5)\.48,6)\.49,7)\.50,8)\.51,9)
\.52,10)\.53,11)\.54,12)\.55,13)\.56,14)\.57,15)\.60,18)\.61,19)\.62,20)
\.63,21)\.64,22)\.65,23)\.66,24)\.68,26)\.69,27)\.70,28)\.71,29)\.72,30)
\.78,36)\.83,41)\.86,44)\.87,45)\.88,46)\.90,48)\.98,56)\.103,61)\.104,62)
\.107,65)\.108,66)\.110,68)\.116,74)\.118,76)\.130,88)\.152,110)\.182,140)
\.58,17)\.64,23)\.41,1)\.42,2)\.43,3)\.44,4)\.45,5)\.47,7)\.48,8)\.49,9)
\.50,10)\.51,11)\.52,12)\.53,13)\.54,14)\.55,15)\.56,16)\.57,17)\.58,18)
\.59,19)\.63,23)\.67,27)\.68,28)\.69,29)\.71,31)\.73,33)\.75,35)\.79,39)
\.81,41)\.85,45)\.89,49)\.93,53)\.46,7)\.47,8)\.48,9)\.50,11)\.59,20)\.60,21)
\.63,24)\.65,26)\.66,27)\.74,35)\.76,37)\.96,57)\.107,68)\.46,8)\.56,18)
\.57,19)\.64,26)\.66,28)\.75,37)\.81,43)\.90,52)\.48,11)\.58,21)\.107,70)
\.36,0)\.37,1)\.38,2)\.39,3)\.40,4)\.41,5)\.42,6)\.43,7)\.44,8)\.45,9)\.46,10)
\.47,11)\.48,12)\.49,13)\.50,14)\.51,15)\.52,16)\.53,17)\.55,19)\.56,20)
\.57,21)\.58,22)\.59,23)\.60,24)\.61,25)\.62,26)\.63,27)\.64,28)\.65,29)
\.66,30)\.67,31)\.68,32)\.69,33)\.70,34)\.72,36)\.74,38)\.76,40)\.77,41)
\.78,42)\.80,44)\.81,45)\.82,46)\.84,48)\.86,50)\.88,52)\.89,53)\.90,54)
\.92,56)\.93,57)\.97,61)\.98,62)\.100,64)\.105,69)\.109,73)\.117,81)\.124,88)
\.43,8)\.44,9)\.49,14)\.54,19)\.38,4)\.44,10)\.46,12)\.55,21)\.60,26)\.72,38)
\.80,46)\.39,6)\.41,8)\.42,9)\.49,16)\.50,17)\.51,18)\.54,21)\.56,23)\.60,27)
\.61,28)\.65,32)\.85,52)\.35,3)\.36,4)\.37,5)\.39,7)\.40,8)\.41,9)\.42,10)
\.43,11)\.46,14)\.47,15)\.48,16)\.49,17)\.50,18)\.51,19)\.52,20)\.53,21)
\.54,22)\.55,23)\.56,24)\.57,25)\.59,27)\.61,29)\.64,32)\.67,35)\.69,37)
\.71,39)\.75,43)\.78,46)\.79,47)\.83,51)\.99,67)\.36,5)\.52,21)\.60,29)
\.68,37)\.89,58)\.30,0)\.31,1)\.32,2)\.33,3)\.34,4)\.35,5)\.36,6)\.37,7)
\.38,8)\.39,9)\.40,10)\.41,11)\.42,12)\.43,13)\.44,14)\.45,15)\.46,16)\.47,17)
\.48,18)\.49,19)\.50,20)\.52,22)\.53,23)\.54,24)\.56,26)\.57,27)\.59,29)
\.60,30)\.62,32)\.63,33)\.64,34)\.67,37)\.68,38)\.69,39)\.70,40)\.71,41)
\.76,46)\.77,47)\.78,48)\.81,51)\.84,54)\.87,57)\.98,68)\.104,74)\.113,83)
\.124,94)\.126,96)\.132,102)\.194,164)\.252,222)\.36,7)\.39,10)\.79,50)
\.29,1)\.33,5)\.35,7)\.36,8)\.37,9)\.38,10)\.39,11)\.40,12)\.41,13)\.42,14)
\.44,16)\.45,17)\.47,19)\.48,20)\.50,22)\.51,23)\.55,27)\.57,29)\.58,30)
\.61,33)\.62,34)\.63,35)\.67,39)\.71,43)\.72,44)\.77,49)\.81,53)\.89,61)
\.97,69)\.104,76)\.121,93)\.32,5)\.34,7)\.35,8)\.37,10)\.38,11)\.41,14)
\.42,15)\.43,16)\.44,17)\.45,18)\.46,19)\.49,22)\.50,23)\.52,25)\.53,26)
\.55,28)\.56,29)\.57,30)\.62,35)\.65,38)\.72,45)\.74,47)\.80,53)\.87,60)
\.92,65)\.98,71)\.28,2)\.34,8)\.35,9)\.44,18)\.45,19)\.47,21)\.49,23)\.52,26)
\.57,31)\.67,41)\.71,45)\.39,14)\.49,24)\.25,1)\.26,2)\.27,3)\.28,4)\.29,5)
\.30,6)\.31,7)\.32,8)\.33,9)\.34,10)\.35,11)\.36,12)\.37,13)\.38,14)\.39,15)
\.40,16)\.41,17)\.42,18)\.43,19)\.44,20)\.45,21)\.46,22)\.47,23)\.48,24)
\.49,25)\.50,26)\.51,27)\.52,28)\.53,29)\.54,30)\.55,31)\.56,32)\.57,33)
\.58,34)\.59,35)\.60,36)\.61,37)\.62,38)\.63,39)\.64,40)\.65,41)\.66,42)
\.67,43)\.68,44)\.71,47)\.73,49)\.74,50)\.79,55)\.81,57)\.83,59)\.85,61)
\.91,67)\.95,71)\.96,72)\.101,77)\.113,89)\.127,103)\.145,121)\.31,8)\.34,11)
\.38,15)\.42,19)\.88,65)\.30,8)\.33,11)\.34,12)\.36,14)\.39,17)\.42,20)
\.43,21)\.50,28)\.51,29)\.57,35)\.78,56)\.28,7)\.32,11)\.34,13)\.35,14)
\.37,16)\.38,17)\.41,20)\.43,22)\.44,23)\.46,25)\.47,26)\.50,29)\.51,30)
\.52,31)\.53,32)\.56,35)\.59,38)\.68,47)\.79,58)\.94,73)\.21,1)\.25,5)\.26,6)
\.27,7)\.28,8)\.29,9)\.30,10)\.31,11)\.32,12)\.33,13)\.34,14)\.35,15)\.36,16)
\.37,17)\.38,18)\.39,19)\.41,21)\.44,24)\.45,25)\.47,27)\.49,29)\.52,32)
\.53,33)\.55,35)\.56,36)\.65,45)\.68,48)\.69,49)\.70,50)\.75,55)\.79,59)
\.84,64)\.90,70)\.30,11)\.32,13)\.35,16)\.20,2)\.21,3)\.22,4)\.23,5)\.24,6)
\.25,7)\.26,8)\.27,9)\.28,10)\.29,11)\.30,12)\.31,13)\.32,14)\.33,15)\.34,16)
\.35,17)\.36,18)\.37,19)\.38,20)\.39,21)\.40,22)\.41,23)\.42,24)\.43,25)
\.44,26)\.45,27)\.46,28)\.47,29)\.48,30)\.49,31)\.50,32)\.51,33)\.52,34)
\.53,35)\.56,38)\.59,41)\.60,42)\.61,43)\.62,44)\.63,45)\.65,47)\.68,50)
\.69,51)\.72,54)\.74,56)\.76,58)\.80,62)\.81,63)\.87,69)\.90,72)\.95,77)
\.101,83)\.110,92)\.115,97)\.116,98)\.120,102)\.166,148)\.19,3)\.20,4)\.22,6)
\.23,7)\.24,8)\.25,9)\.26,10)\.27,11)\.29,13)\.30,14)\.31,15)\.32,16)\.33,17)
\.34,18)\.35,19)\.36,20)\.37,21)\.38,22)\.39,23)\.40,24)\.41,25)\.43,27)
\.44,28)\.45,29)\.46,30)\.47,31)\.51,35)\.54,38)\.55,39)\.56,40)\.60,44)
\.67,51)\.103,87)\.22,7)\.24,9)\.25,10)\.26,11)\.27,12)\.28,13)\.29,14)
\.30,15)\.32,17)\.33,18)\.34,19)\.36,21)\.38,23)\.39,24)\.42,27)\.43,28)
\.44,29)\.46,31)\.47,32)\.49,34)\.53,38)\.54,39)\.74,59)\.89,74)\.25,11)
\.30,16)\.31,17)\.34,20)\.37,23)\.38,24)\.40,26)\.41,27)\.43,29)\.100,86)
\.38,25)\.51,38)\.12,0)\.16,4)\.17,5)\.18,6)\.19,7)\.20,8)\.21,9)\.22,10)
\.23,11)\.24,12)\.25,13)\.26,14)\.27,15)\.28,16)\.29,17)\.30,18)\.31,19)
\.32,20)\.33,21)\.34,22)\.35,23)\.36,24)\.37,25)\.38,26)\.39,27)\.40,28)
\.41,29)\.42,30)\.43,31)\.44,32)\.45,33)\.46,34)\.47,35)\.48,36)\.49,37)
\.50,38)\.51,39)\.52,40)\.53,41)\.54,42)\.55,43)\.56,44)\.57,45)\.58,46)
\.62,50)\.63,51)\.64,52)\.66,54)\.70,58)\.71,59)\.72,60)\.73,61)\.77,65)
\.78,66)\.80,68)\.86,74)\.87,75)\.88,76)\.89,77)\.92,80)\.98,86)\.122,110)
\.26,15)\.28,17)\.37,26)\.18,8)\.20,10)\.24,14)\.25,15)\.27,17)\.29,19)
\.31,21)\.33,23)\.34,24)\.37,27)\.39,29)\.40,30)\.42,32)\.55,45)\.59,49)
\.62,52)\.63,53)\.74,64)\.104,94)\.13,4)\.16,7)\.17,8)\.19,10)\.20,11)\.21,12)
\.22,13)\.23,14)\.24,15)\.25,16)\.28,19)\.29,20)\.30,21)\.32,23)\.33,24)
\.35,26)\.36,27)\.37,28)\.40,31)\.44,35)\.46,37)\.47,38)\.50,41)\.53,44)
\.55,46)\.56,47)\.57,48)\.59,50)\.62,53)\.79,70)\.86,77)\.15,7)\.17,9)\.18,10)
\.19,11)\.20,12)\.21,13)\.22,14)\.23,15)\.24,16)\.25,17)\.26,18)\.27,19)
\.28,20)\.29,21)\.30,22)\.31,23)\.32,24)\.33,25)\.34,26)\.35,27)\.36,28)
\.37,29)\.39,31)\.40,32)\.41,33)\.42,34)\.43,35)\.45,37)\.47,39)\.50,42)
\.51,43)\.52,44)\.58,50)\.61,53)\.62,54)\.79,71)\.22,15)\.26,19)\.27,20)
\.34,27)\.52,45)\.72,65)\.12,6)\.13,7)\.15,9)\.16,10)\.17,11)\.18,12)\.19,13)
\.20,14)\.21,15)\.22,16)\.23,17)\.24,18)\.25,19)\.26,20)\.27,21)\.28,22)
\.29,23)\.30,24)\.31,25)\.32,26)\.33,27)\.34,28)\.35,29)\.36,30)\.37,31)
\.38,32)\.39,33)\.40,34)\.41,35)\.42,36)\.43,37)\.44,38)\.46,40)\.47,41)
\.48,42)\.49,43)\.50,44)\.51,45)\.53,47)\.54,48)\.55,49)\.57,51)\.61,55)
\.67,61)\.68,62)\.70,64)\.76,70)\.81,75)\.86,80)\.24,19)\.28,23)\.29,24)
\.34,29)\.39,34)\.46,41)\.14,10)\.16,12)\.17,13)\.18,14)\.19,15)\.20,16)
\.21,17)\.22,18)\.24,20)\.25,21)\.27,23)\.28,24)\.29,25)\.30,26)\.31,27)
\.32,28)\.33,29)\.35,31)\.36,32)\.37,33)\.38,34)\.44,40)\.47,43)\.48,44)
\.50,46)\.51,47)\.53,49)\.60,56)\.61,57)\.14,11)\.16,13)\.17,14)\.18,15)
\.20,17)\.21,18)\.23,20)\.24,21)\.25,22)\.26,23)\.29,26)\.30,27)\.31,28)
\.32,29)\.34,31)\.35,32)\.37,34)\.38,35)\.40,37)\.43,40)\.45,42)\.47,44)
\.49,46)\.50,47)\.51,48)\.54,51)\.60,57)\.66,63)\.70,67)\.14,12)\.16,14)
\.20,18)\.21,19)\.25,23)\.26,24)\.28,26)\.29,27)\.31,29)\.33,31)\.34,32)
\.37,35)\.38,36)\.42,40)\.47,45)\.49,47)\.56,54)\.79,77)\.22,21)\.33,32)
\.46,45)\.3,3)\.7,7)\.9,9)\.10,10)\.11,11)\.12,12)\.13,13)\.14,14)\.15,15)
\.16,16)\.17,17)\.18,18)\.19,19)\.20,20)\.21,21)\.22,22)\.23,23)\.24,24)
\.25,25)\.26,26)\.27,27)\.28,28)\.29,29)\.30,30)\.31,31)\.32,32)\.33,33)
\.34,34)\.35,35)\.36,36)\.37,37)\.38,38)\.39,39)\.40,40)\.41,41)\.43,43)
\.44,44)\.45,45)\.46,46)\.47,47)\.48,48)\.49,49)\.52,52)\.53,53)\.55,55)
\.56,56)\.57,57)\.59,59)\.61,61)\.62,62)\.63,63)\.65,65)\.67,67)\.69,69)
\.71,71)\.75,75)\.77,77)\.78,78)\.79,79)\.81,81)\.83,83)\.85,85)\.87,87)
\.89,89)\.91,91)\.95,95)\.97,97)\.103,103)\.107,107)\.111,111)\.119,119)
\.121,121)\.123,123)\.131,131)\.143,143)\.149,149)\.151,151)\.179,179)
\.223,223)
\.251,251)\.12,13)\.16,17)\.20,21)\.22,23)\.34,35)\.9,11)\.15,17)\.18,20)
\.20,22)\.21,23)\.24,26)\.25,27)\.26,28)\.27,29)\.29,31)\.32,34)\.33,35)
\.39,41)\.40,42)\.45,47)\.50,52)\.13,16)\.14,17)\.15,18)\.16,19)\.17,20)
\.18,21)\.19,22)\.20,23)\.21,24)\.23,26)\.26,29)\.27,30)\.29,32)\.31,34)
\.32,35)\.34,37)\.36,39)\.37,40)\.40,43)\.42,45)\.46,49)\.47,50)\.48,51)
\.67,70)\.74,77)\.11,15)\.13,17)\.14,18)\.15,19)\.16,20)\.17,21)\.18,22)
\.19,23)\.20,24)\.21,25)\.22,26)\.23,27)\.24,28)\.25,29)\.26,30)\.28,32)
\.29,33)\.31,35)\.32,36)\.33,37)\.37,41)\.39,43)\.40,44)\.43,47)\.49,53)
\.55,59)\.56,60)\.17,22)\.20,25)\.22,27)\.24,29)\.34,39)\.39,44)\.52,57)
\.68,73)\.70,75)\.6,12)\.7,13)\.8,14)\.9,15)\.10,16)\.11,17)\.12,18)\.13,19)
\.14,20)\.15,21)\.16,22)\.17,23)\.18,24)\.19,25)\.20,26)\.21,27)\.22,28)
\.23,29)\.24,30)\.25,31)\.26,32)\.27,33)\.28,34)\.29,35)\.30,36)\.32,38)
\.33,39)\.35,41)\.36,42)\.37,43)\.38,44)\.40,46)\.41,47)\.42,48)\.44,50)
\.45,51)\.46,52)\.47,53)\.48,54)\.49,55)\.51,57)\.55,61)\.58,64)\.61,67)
\.62,68)\.70,76)\.75,81)\.19,26)\.20,27)\.26,33)\.27,34)\.8,16)\.9,17)\.10,18)
\.11,19)\.13,21)\.14,22)\.15,23)\.16,24)\.17,25)\.18,26)\.19,27)\.20,28)
\.21,29)\.22,30)\.23,31)\.24,32)\.25,33)\.26,34)\.27,35)\.28,36)\.29,37)
\.30,38)\.31,39)\.32,40)\.33,41)\.34,42)\.35,43)\.37,45)\.38,46)\.39,47)
\.42,50)\.43,51)\.44,52)\.45,53)\.53,61)\.54,62)\.57,65)\.6,15)\.8,17)\.10,19)
\.11,20)\.12,21)\.13,22)\.14,23)\.16,25)\.17,26)\.18,27)\.19,28)\.20,29)
\.22,31)\.23,32)\.24,33)\.26,35)\.27,36)\.29,38)\.30,39)\.31,40)\.32,41)
\.35,44)\.36,45)\.38,47)\.40,49)\.41,50)\.44,53)\.46,55)\.47,56)\.48,57)
\.50,59)\.70,79)\.15,25)\.18,28)\.19,29)\.22,32)\.24,34)\.25,35)\.27,37)
\.29,39)\.30,40)\.37,47)\.40,50)\.49,59)\.52,62)\.94,104)\.112,122)\.15,26)
\.30,41)\.0,12)\.4,16)\.5,17)\.6,18)\.7,19)\.8,20)\.9,21)\.10,22)\.11,23)
\.12,24)\.13,25)\.14,26)\.15,27)\.16,28)\.17,29)\.18,30)\.19,31)\.20,32)
\.21,33)\.22,34)\.23,35)\.24,36)\.25,37)\.26,38)\.27,39)\.28,40)\.29,41)
\.30,42)\.31,43)\.32,44)\.33,45)\.34,46)\.35,47)\.36,48)\.37,49)\.38,50)
\.39,51)\.40,52)\.41,53)\.42,54)\.43,55)\.44,56)\.45,57)\.46,58)\.47,59)
\.50,62)\.51,63)\.52,64)\.54,66)\.58,70)\.59,71)\.60,72)\.61,73)\.65,77)
\.68,80)\.74,86)\.76,88)\.77,89)\.80,92)\.86,98)\.110,122)\.30,43)\.31,44)
\.12,26)\.20,34)\.22,36)\.24,38)\.25,39)\.26,40)\.27,41)\.29,43)\.32,46)
\.36,50)\.42,56)\.66,80)\.9,24)\.10,25)\.12,27)\.14,29)\.17,32)\.18,33)
\.19,34)\.20,35)\.21,36)\.23,38)\.24,39)\.29,44)\.31,46)\.32,47)\.33,48)
\.34,49)\.39,54)\.44,59)\.53,68)\.59,74)\.74,89)\.6,22)\.7,23)\.9,25)\.10,26)
\.11,27)\.12,28)\.13,29)\.14,30)\.15,31)\.17,33)\.18,34)\.19,35)\.20,36)
\.21,37)\.22,38)\.23,39)\.24,40)\.25,41)\.27,43)\.28,44)\.29,45)\.30,46)
\.31,47)\.33,49)\.34,50)\.35,51)\.38,54)\.39,55)\.40,56)\.41,57)\.44,60)
\.47,63)\.49,65)\.51,67)\.55,71)\.58,74)\.87,103)\.14,31)\.23,40)\.26,43)
\.33,50)\.60,77)\.3,21)\.4,22)\.5,23)\.6,24)\.7,25)\.8,26)\.9,27)\.10,28)
\.11,29)\.12,30)\.13,31)\.14,32)\.15,33)\.16,34)\.17,35)\.18,36)\.19,37)
\.20,38)\.21,39)\.22,40)\.23,41)\.24,42)\.25,43)\.26,44)\.27,45)\.28,46)
\.29,47)\.30,48)\.31,49)\.32,50)\.33,51)\.34,52)\.35,53)\.38,56)\.40,58)
\.41,59)\.42,60)\.44,62)\.45,63)\.46,64)\.47,65)\.48,66)\.50,68)\.54,72)
\.56,74)\.58,76)\.62,80)\.63,81)\.69,87)\.72,90)\.77,95)\.83,101)\.92,110)
\.97,115)\.98,116)\.102,120)\.148,166)\.9,28)\.16,35)\.25,44)\.26,45)\.1,21)
\.5,25)\.7,27)\.8,28)\.9,29)\.12,32)\.13,33)\.14,34)\.15,35)\.16,36)\.17,37)
\.18,38)\.19,39)\.20,40)\.21,41)\.24,44)\.25,45)\.26,46)\.27,47)\.29,49)
\.30,50)\.31,51)\.33,53)\.35,55)\.37,57)\.39,59)\.43,63)\.45,65)\.48,68)
\.49,69)\.50,70)\.55,75)\.64,84)\.7,28)\.9,30)\.11,32)\.12,33)\.14,35)\.16,37)
\.17,38)\.18,39)\.20,41)\.21,42)\.23,44)\.26,47)\.29,50)\.30,51)\.34,55)
\.35,56)\.36,57)\.38,59)\.47,68)\.11,33)\.13,35)\.14,36)\.15,37)\.17,39)
\.18,40)\.19,41)\.20,42)\.21,43)\.22,44)\.26,48)\.28,50)\.29,51)\.33,55)
\.36,58)\.37,59)\.43,65)\.1,25)\.2,26)\.3,27)\.4,28)\.5,29)\.6,30)\.7,31)
\.8,32)\.9,33)\.10,34)\.11,35)\.12,36)\.13,37)\.14,38)\.15,39)\.16,40)\.17,41)
\.18,42)\.19,43)\.20,44)\.21,45)\.22,46)\.23,47)\.24,48)\.25,49)\.26,50)
\.27,51)\.28,52)\.29,53)\.30,54)\.31,55)\.32,56)\.33,57)\.34,58)\.35,59)
\.36,60)\.37,61)\.38,62)\.39,63)\.40,64)\.41,65)\.42,66)\.43,67)\.44,68)
\.45,69)\.47,71)\.49,73)\.50,74)\.52,76)\.55,79)\.57,81)\.59,83)\.61,85)
\.67,91)\.71,95)\.77,101)\.89,113)\.103,127)\.121,145)\.10,35)\.14,39)\.19,44)
\.22,47)\.24,49)\.34,59)\.6,32)\.12,38)\.15,41)\.19,45)\.21,47)\.28,54)
\.41,67)\.45,71)\.7,34)\.8,35)\.10,37)\.11,38)\.12,39)\.14,41)\.15,42)\.16,43)
\.17,44)\.22,49)\.23,50)\.25,52)\.26,53)\.28,55)\.29,56)\.30,57)\.34,61)
\.35,62)\.37,64)\.38,65)\.40,67)\.45,72)\.53,80)\.56,83)\.60,87)\.65,92)
\.5,33)\.8,36)\.9,37)\.10,38)\.11,39)\.12,40)\.13,41)\.15,43)\.16,44)\.17,45)
\.19,47)\.20,48)\.21,49)\.22,50)\.23,51)\.25,53)\.26,54)\.28,56)\.29,57)
\.30,58)\.31,59)\.38,66)\.39,67)\.40,68)\.43,71)\.46,74)\.48,76)\.49,77)
\.61,89)\.69,97)\.76,104)\.93,121)\.28,57)\.30,59)\.0,30)\.1,31)\.2,32)
\.3,33)\.5,35)\.6,36)\.7,37)\.8,38)\.9,39)\.10,40)\.11,41)\.12,42)\.13,43)
\.14,44)\.15,45)\.16,46)\.17,47)\.18,48)\.19,49)\.20,50)\.22,52)\.23,53)
\.24,54)\.26,56)\.27,57)\.29,59)\.30,60)\.32,62)\.33,63)\.34,64)\.35,65)
\.38,68)\.40,70)\.41,71)\.42,72)\.45,75)\.46,76)\.47,77)\.51,81)\.54,84)
\.57,87)\.68,98)\.74,104)\.83,113)\.94,124)\.96,126)\.102,132)\.164,194)
\.222,252)\.15,46)\.16,47)\.22,53)\.37,68)\.3,35)\.5,37)\.7,39)\.8,40)\.9,41)
\.11,43)\.12,44)\.13,45)\.14,46)\.15,47)\.16,48)\.17,49)\.18,50)\.19,51)
\.22,54)\.23,55)\.24,56)\.25,57)\.26,58)\.27,59)\.29,61)\.32,64)\.34,66)
\.35,67)\.37,69)\.39,71)\.51,83)\.61,93)\.67,99)\.4,37)\.8,41)\.10,43)\.12,45)
\.14,47)\.15,48)\.16,49)\.18,51)\.20,53)\.21,54)\.23,56)\.26,59)\.27,60)
\.28,61)\.29,62)\.31,64)\.32,65)\.48,81)\.52,85)\.63,96)\.4,38)\.10,44)
\.13,47)\.15,49)\.16,50)\.17,51)\.22,56)\.27,61)\.38,72)\.58,92)\.9,44)
\.14,49)\.19,54)\.20,55)\.27,62)\.30,65)\.33,68)\.50,85)\.0,36)\.1,37)\.2,38)
\.3,39)\.4,40)\.5,41)\.6,42)\.7,43)\.8,44)\.9,45)\.10,46)\.11,47)\.12,48)
\.13,49)\.14,50)\.15,51)\.16,52)\.17,53)\.18,54)\.19,55)\.20,56)\.21,57)
\.22,58)\.23,59)\.24,60)\.25,61)\.26,62)\.27,63)\.28,64)\.29,65)\.30,66)
\.31,67)\.32,68)\.33,69)\.34,70)\.36,72)\.37,73)\.38,74)\.40,76)\.41,77)
\.42,78)\.43,79)\.44,80)\.45,81)\.46,82)\.48,84)\.50,86)\.52,88)\.53,89)
\.54,90)\.56,92)\.57,93)\.61,97)\.62,98)\.64,100)\.73,109)\.81,117)\.88,124)
\.18,55)\.22,59)\.30,67)\.42,79)\.8,46)\.12,50)\.15,53)\.19,57)\.28,66)
\.31,69)\.36,74)\.37,75)\.43,81)\.8,47)\.11,50)\.18,57)\.20,59)\.22,61)
\.25,64)\.26,65)\.27,66)\.28,67)\.35,74)\.36,75)\.57,96)\.62,101)\.3,43)
\.4,44)\.7,47)\.9,49)\.10,50)\.11,51)\.12,52)\.13,53)\.14,54)\.15,55)\.16,56)
\.17,57)\.18,58)\.19,59)\.23,63)\.25,65)\.27,67)\.28,68)\.29,69)\.30,70)
\.31,71)\.33,73)\.39,79)\.40,80)\.0,42)\.3,45)\.4,46)\.5,47)\.6,48)\.7,49)
\.8,50)\.9,51)\.10,52)\.11,53)\.12,54)\.13,55)\.14,56)\.15,57)\.16,58)\.18,60)
\.19,61)\.20,62)\.21,63)\.22,64)\.23,65)\.24,66)\.25,67)\.26,68)\.27,69)
\.28,70)\.30,72)\.32,74)\.35,77)\.40,82)\.41,83)\.44,86)\.46,88)\.48,90)
\.56,98)\.60,102)\.61,103)\.62,104)\.65,107)\.66,108)\.68,110)\.74,116)
\.76,118)\.88,130)\.110,152)\.140,182)\.7,50)\.17,60)\.18,61)\.22,65)\.42,85)
\.5,49)\.8,52)\.9,53)\.10,54)\.11,55)\.12,56)\.13,57)\.15,59)\.19,63)\.20,64)
\.22,66)\.29,73)\.30,74)\.31,75)\.42,86)\.57,101)\.4,49)\.8,53)\.9,54)\.10,55)
\.11,56)\.13,58)\.14,59)\.17,62)\.19,64)\.22,67)\.24,69)\.25,70)\.26,71)
\.28,73)\.30,75)\.32,77)\.35,80)\.43,88)\.44,89)\.48,93)\.59,104)\.63,108)
\.7,53)\.9,55)\.10,56)\.15,61)\.16,62)\.24,70)\.29,75)\.35,81)\.52,98)\.0,48)
\.1,49)\.3,51)\.4,52)\.5,53)\.6,54)\.7,55)\.8,56)\.9,57)\.10,58)\.11,59)
\.12,60)\.13,61)\.14,62)\.15,63)\.16,64)\.17,65)\.18,66)\.19,67)\.20,68)
\.21,69)\.22,70)\.23,71)\.24,72)\.25,73)\.27,75)\.28,76)\.29,77)\.30,78)
\.31,79)\.32,80)\.33,81)\.35,83)\.36,84)\.37,85)\.38,86)\.39,87)\.43,91)
\.45,93)\.47,95)\.48,96)\.49,97)\.53,101)\.55,103)\.59,107)\.69,117)\.71,119)
\.78,126)\.82,130)\.101,149)\.119,167)\.147,195)\.177,225)\.6,56)\.8,58)
\.9,59)\.10,60)\.14,64)\.15,65)\.19,69)\.23,73)\.24,74)\.34,84)\.59,109)
\.5,56)\.8,59)\.12,63)\.15,66)\.16,67)\.18,69)\.21,72)\.23,74)\.24,75)\.32,83)
\.42,93)\.1,53)\.3,55)\.4,56)\.5,57)\.7,59)\.8,60)\.9,61)\.11,63)\.12,64)
\.13,65)\.14,66)\.15,67)\.16,68)\.17,69)\.19,71)\.24,76)\.25,77)\.27,79)
\.28,80)\.33,85)\.36,88)\.40,92)\.51,103)\.8,61)\.38,91)\.0,54)\.1,55)\.2,56)
\.3,57)\.4,58)\.6,60)\.7,61)\.8,62)\.9,63)\.10,64)\.11,65)\.12,66)\.13,67)
\.14,68)\.15,69)\.16,70)\.17,71)\.18,72)\.20,74)\.21,75)\.23,77)\.26,80)
\.27,81)\.29,83)\.30,84)\.32,86)\.33,87)\.34,88)\.38,92)\.39,93)\.48,102)
\.56,110)\.62,116)\.68,122)\.79,133)\.9,64)\.2,58)\.3,59)\.5,61)\.7,63)
\.9,65)\.10,66)\.11,67)\.12,68)\.13,69)\.14,70)\.15,71)\.16,72)\.17,73)
\.18,74)\.19,75)\.20,76)\.23,79)\.26,82)\.27,83)\.31,87)\.34,90)\.35,91)
\.41,97)\.47,103)\.8,65)\.10,67)\.11,68)\.14,71)\.16,73)\.17,74)\.20,77)
\.21,78)\.22,79)\.23,80)\.24,81)\.28,85)\.29,86)\.35,92)\.44,101)\.45,102)
\.56,113)\.7,65)\.11,69)\.13,71)\.25,83)\.33,91)\.71,129)\.13,72)\.1,61)
\.2,62)\.3,63)\.4,64)\.5,65)\.6,66)\.7,67)\.8,68)\.9,69)\.10,70)\.11,71)
\.12,72)\.13,73)\.14,74)\.15,75)\.16,76)\.17,77)\.18,78)\.19,79)\.20,80)
\.21,81)\.22,82)\.23,83)\.24,84)\.25,85)\.26,86)\.27,87)\.28,88)\.29,89)
\.31,91)\.32,92)\.34,94)\.35,95)\.36,96)\.37,97)\.38,98)\.39,99)\.40,100)
\.41,101)\.43,103)\.44,104)\.45,105)\.46,106)\.47,107)\.48,108)\.49,109)
\.54,114)\.59,119)\.65,125)\.68,128)\.76,136)\.94,154)\.108,168)\.109,169)
\.26,87)\.6,68)\.16,78)\.22,84)\.50,112)\.3,66)\.7,70)\.8,71)\.11,74)\.12,75)
\.13,76)\.14,77)\.18,81)\.19,82)\.20,83)\.23,86)\.26,89)\.27,90)\.35,98)
\.40,103)\.5,69)\.6,70)\.7,71)\.8,72)\.9,73)\.11,75)\.13,77)\.14,78)\.15,79)
\.19,83)\.21,85)\.22,86)\.23,87)\.24,88)\.25,89)\.27,91)\.28,92)\.29,93)
\.30,94)\.31,95)\.37,101)\.39,103)\.6,71)\.10,75)\.14,79)\.22,87)\.3,69)
\.4,70)\.5,71)\.6,72)\.7,73)\.9,75)\.10,76)\.12,78)\.13,79)\.14,80)\.15,81)
\.17,83)\.18,84)\.19,85)\.20,86)\.24,90)\.27,93)\.30,96)\.31,97)\.42,108)
\.53,119)\.56,122)\.68,134)\.70,136)\.86,152)\.118,184)\.13,80)\.5,73)\.7,75)
\.8,76)\.9,77)\.11,79)\.12,80)\.13,81)\.14,82)\.17,85)\.18,86)\.21,89)\.22,90)
\.23,91)\.29,97)\.34,102)\.35,103)\.57,125)\.5,74)\.8,77)\.9,78)\.14,83)
\.17,86)\.28,97)\.41,110)\.53,122)\.4,74)\.5,75)\.7,77)\.9,79)\.11,81)\.12,82)
\.13,83)\.14,84)\.19,89)\.22,92)\.25,95)\.30,100)\.34,104)\.39,109)\.44,114)
\.29,100)\.1,73)\.2,74)\.3,75)\.4,76)\.5,77)\.6,78)\.7,79)\.8,80)\.9,81)
\.10,82)\.11,83)\.12,84)\.13,85)\.14,86)\.15,87)\.16,88)\.17,89)\.18,90)
\.19,91)\.20,92)\.21,93)\.23,95)\.24,96)\.25,97)\.26,98)\.27,99)\.29,101)
\.31,103)\.33,105)\.34,106)\.38,110)\.39,111)\.40,112)\.41,113)\.43,115)
\.46,118)\.50,122)\.51,123)\.55,127)\.58,130)\.59,131)\.67,139)\.71,143)
\.32,105)\.60,133)\.9,83)\.12,86)\.16,90)\.8,83)\.9,84)\.11,86)\.12,87)
\.13,88)\.16,91)\.17,92)\.24,99)\.32,107)\.38,113)\.39,114)\.53,128)\.1,77)
\.4,80)\.5,81)\.10,86)\.13,89)\.16,92)\.17,93)\.21,97)\.27,103)\.29,105)
\.36,112)\.6,83)\.10,87)\.4,82)\.5,83)\.7,85)\.8,86)\.11,89)\.12,90)\.13,91)
\.14,92)\.15,93)\.16,94)\.17,95)\.18,96)\.21,99)\.22,100)\.23,101)\.24,102)
\.27,105)\.35,113)\.38,116)\.50,128)\.66,144)\.77,155)\.78,156)\.176,254)
\.3,83)\.5,85)\.7,87)\.8,88)\.9,89)\.11,91)\.14,94)\.15,95)\.17,97)\.20,100)
\.22,102)\.23,103)\.27,107)\.35,115)\.38,118)\.45,125)\.49,129)\.8,89)\.16,97)
\.17,98)\.20,101)\.23,104)\.25,106)\.26,107)\.6,88)\.12,94)\.0,84)\.1,85)
\.2,86)\.3,87)\.5,89)\.6,90)\.7,91)\.8,92)\.9,93)\.11,95)\.12,96)\.13,97)
\.14,98)\.15,99)\.16,100)\.17,101)\.18,102)\.19,103)\.20,104)\.21,105)\.22,106)
\.25,109)\.27,111)\.28,112)\.29,113)\.30,114)\.32,116)\.34,118)\.40,124)
\.42,126)\.45,129)\.47,131)\.50,134)\.86,170)\.4,89)\.16,101)\.9,95)\.14,100)
\.18,104)\.5,92)\.28,115)\.35,122)\.40,127)\.50,137)\.2,90)\.5,93)\.6,94)
\.7,95)\.8,96)\.9,97)\.10,98)\.13,101)\.19,107)\.20,108)\.22,110)\.43,131)
\.51,139)\.12,101)\.24,113)\.0,90)\.4,94)\.5,95)\.7,97)\.8,98)\.9,99)\.10,100)
\.11,101)\.12,102)\.14,104)\.15,105)\.17,107)\.18,108)\.20,110)\.22,112)
\.23,113)\.24,114)\.26,116)\.27,117)\.29,119)\.32,122)\.34,124)\.39,129)
\.42,132)\.47,137)\.48,138)\.68,158)\.69,159)\.98,188)\.138,228)\.7,98)
\.5,97)\.7,99)\.9,101)\.10,102)\.12,104)\.15,107)\.19,111)\.21,113)\.22,114)
\.28,120)\.44,136)\.2,95)\.10,103)\.11,104)\.15,108)\.17,110)\.37,130)\.3,99)
\.5,101)\.6,102)\.7,103)\.8,104)\.9,105)\.10,106)\.11,107)\.13,109)\.14,110)
\.15,111)\.16,112)\.17,113)\.18,114)\.19,115)\.21,117)\.23,119)\.27,123)
\.29,125)\.33,129)\.35,131)\.37,133)\.41,137)\.44,140)\.47,143)\.51,147)
\.55,151)\.61,157)\.65,161)\.77,173)\.4,102)\.11,109)\.26,124)\.37,135)
\.13,112)\.16,115)\.20,119)\.26,125)\.1,101)\.3,103)\.4,104)\.7,107)\.8,108)
\.9,109)\.13,113)\.14,114)\.17,117)\.18,118)\.22,122)\.24,124)\.25,125)
\.28,128)\.34,134)\.43,143)\.22,123)\.1,103)\.3,105)\.6,108)\.8,110)\.9,111)
\.11,113)\.12,114)\.14,116)\.16,118)\.20,122)\.31,133)\.33,135)\.44,146)
\.2,106)\.7,111)\.11,115)\.17,121)\.4,109)\.6,111)\.14,119)\.18,123)\.19,124)
\.20,125)\.23,128)\.30,135)\.33,138)\.58,163)\.10,116)\.16,122)\.22,128)
\.26,132)\.2,110)\.4,112)\.6,114)\.8,116)\.9,117)\.10,118)\.12,120)\.13,121)
\.14,122)\.15,123)\.16,124)\.17,125)\.18,126)\.21,129)\.22,130)\.23,131)
\.25,133)\.28,136)\.30,138)\.33,141)\.36,144)\.37,145)\.40,148)\.41,149)
\.42,150)\.52,160)\.66,174)\.68,176)\.83,191)\.92,200)\.8,118)\.18,128)
\.24,134)\.54,164)\.11,122)\.12,123)\.46,157)\.5,117)\.6,118)\.7,119)\.11,123)
\.13,125)\.15,127)\.16,128)\.17,129)\.19,131)\.33,145)\.6,120)\.8,122)\.9,123)
\.12,126)\.13,127)\.14,128)\.15,129)\.16,130)\.28,142)\.31,145)\.32,146)
\.41,155)\.44,158)\.116,230)\.7,122)\.5,121)\.9,125)\.19,135)\.21,137)\.7,124)
\.8,125)\.12,129)\.50,167)\.17,135)\.22,140)\.27,145)\.2,122)\.3,123)\.5,125)
\.7,127)\.9,129)\.10,130)\.11,131)\.12,132)\.13,133)\.15,135)\.19,139)\.21,141)
\.23,143)\.27,147)\.29,149)\.34,154)\.43,163)\.49,169)\.53,173)\.56,176)
\.74,194)\.137,257)\.9,130)\.34,155)\.22,145)\.31,154)\.57,180)\.5,129)
\.25,149)\.2,128)\.6,132)\.8,134)\.12,138)\.16,142)\.18,144)\.20,146)\.24,150)
\.26,152)\.38,164)\.40,166)\.76,202)\.3,131)\.14,142)\.15,143)\.16,144)
\.19,147)\.23,151)\.53,181)\.11,140)\.17,146)\.2,132)\.10,140)\.6,138)\.8,140)
\.9,141)\.11,143)\.13,145)\.14,146)\.15,147)\.16,148)\.20,152)\.23,155)
\.26,158)\.28,160)\.41,173)\.48,180)\.65,197)\.18,151)\.44,179)\.6,142)
\.7,143)\.15,151)\.21,157)\.6,144)\.8,146)\.12,150)\.17,155)\.18,156)\.24,162)
\.27,165)\.37,175)\.38,176)\.42,180)\.48,186)\.62,200)\.96,234)\.4,144)
\.5,145)\.13,153)\.28,170)\.17,160)\.1,145)\.4,148)\.7,151)\.9,153)\.11,155)
\.13,157)\.17,161)\.19,163)\.21,165)\.22,166)\.23,167)\.31,175)\.35,179)
\.39,183)\.41,185)\.63,207)\.115,259)\.6,152)\.6,153)\.20,167)\.1,149)\.9,157)
\.13,161)\.9,159)\.14,164)\.15,165)\.22,172)\.24,174)\.34,184)\.38,188)
\.40,190)\.12,164)\.8,161)\.12,165)\.32,185)\.5,161)\.8,164)\.9,165)\.11,167)
\.12,168)\.17,173)\.18,174)\.20,176)\.34,190)\.81,237)\.52,210)\.19,178)
\.5,165)\.7,167)\.15,175)\.19,179)\.23,183)\.31,191)\.16,177)\.3,165)\.10,172)
\.11,173)\.30,192)\.34,196)\.35,197)\.50,212)\.8,173)\.28,193)\.48,213)
\.5,173)\.10,178)\.13,181)\.15,183)\.17,185)\.19,187)\.27,195)\.31,199)
\.72,240)\.95,263)\.14,184)\.31,202)\.6,180)\.12,186)\.26,200)\.32,206)
\.7,183)\.41,218)\.12,190)\.21,199)\.5,185)\.6,186)\.7,187)\.13,193)\.16,196)
\.24,204)\.39,219)\.63,243)\.6,188)\.10,194)\.4,190)\.8,194)\.20,206)\.36,222)
\.38,224)\.60,246)\.80,266)\.13,201)\.4,194)\.6,198)\.13,205)\.15,207)\.20,212)
\.21,213)\.25,217)\.27,219)\.29,221)\.35,227)\.8,206)\.12,210)\.32,230)
\.71,269)\.4,208)\.10,214)\.14,218)\.18,222)\.32,236)\.4,214)\.10,220)\.19,229)
\.20,230)\.48,258)\.62,272)\.26,239)\.11,227)\.13,229)\.25,241)\.29,245)
\.59,275)\.6,228)\.39,264)\.3,231)\.10,238)\.14,242)\.17,245)\.18,246)\.22,250)
\.37,265)\.26,260)\.36,270)\.16,254)\.3,243)\.11,251)\.19,259)\.23,263)
\.33,273)\.47,287)\.5,251)\.12,264)\.30,282)\.38,293)\.15,271)\.22,280)
\.36,294)\.7,271)\.11,275)\.27,291)\.35,299)\.2,272)\.14,284)\.32,302)\.15,291)
\.20,302)\.24,306)\.29,311)\.7,295)\.13,301)\.26,320)\.12,318)\.20,326)
\.9,321)\.23,335)\.12,330)\.17,341)\.18,348)\.19,355)\.17,377)\.10,376)
\.15,387)\.14,416)\.13,433)\.12,462)\.11,491)\.3,3)\.5,5)\.9,9)\.13,13)
\.21,21)\.9,9) % #spec=3837
}
\end{document}